\documentclass[a4paper,12pt]{article}
\usepackage[top=3cm, bottom=3cm, width=160mm,bindingoffset=0mm]{geometry} 
\usepackage[french,english]{babel} 
\usepackage[utf8]{inputenc}                                             
\usepackage[T1]{fontenc}                                                 
\usepackage{lmodern}

\usepackage{lipsum}                                                      

\usepackage{soul}

\usepackage{pdfpages}

\usepackage{verbatim}
\usepackage{epigraph}
\usepackage{fncychap}

\usepackage{fancyhdr}                                   
\usepackage{tabularx}
\newcolumntype{Y}{>{\centering\arraybackslash}X}

\newcolumntype{Z}{>{\centering\arraybackslash}X}
\usepackage{multirow}
\usepackage[activate={true,nocompatibility},stretch=10,shrink=10]{microtype}

\usepackage{ragged2e}
\usepackage{everysel}
\usepackage{footmisc}

\usepackage{tikz,pstricks}  
\usepackage{pdftricks}
\usetikzlibrary{shapes}
\usepackage{svg}

\usepackage{amsmath}                            
\usepackage{amssymb}                            
\usepackage{tensor}
\usepackage{mathrsfs}

\usepackage{graphicx}
\usepackage{caption}
\usepackage{subcaption}

\DeclareCaptionSubType*{figure}

\usepackage{array}                                            
\usepackage{xcolor}                                           
\usepackage{soulutf8}
\usepackage{colortbl}                                         

\usepackage{xspace}
\usepackage{titlesec}

\usepackage{minitoc}
\usepackage{hyperref}
\hypersetup{
    unicode,
    colorlinks=true,
    linkcolor={blue!50!black},
    citecolor=magenta,
    filecolor=gray,      
    urlcolor=cyan,
    bookmarks=true,
    bookmarksnumbered=true,
    bookmarksopen=true
}

\usepackage{bookmark}

\usepackage{fancybox}                                  
\usepackage[framemethod=tikz]{mdframed}                

\usepackage{floatrow}

\usepackage{dashbox} 

\usepackage{xifthen}
\usepackage{setspace}

\numberwithin{equation}{section}
\newcounter{subequation}

\newcommand{\subtagarray}[1]{
\setcounter{subequation}{#1}\let\OldTheEquation\theequation
\renewcommand{\theequation}{\arabic{chapter}.\arabic{equation}.\alph{subequation}}
\mbox{}\refstepcounter{equation}
{\renewcommand{\theequation}{\alph{subequation}}
$(\theequation)$}
\addtocounter{equation}{-1}
\let\theequation\OldTheEquation
}

\makeatletter
\newcommand*{\inlineequation}[2][]{%
  \begingroup
    \refstepcounter{equation}%
    \ifx\\#1\\%
    \else
      \label{#1}%
    \fi
    \relpenalty=10000 %
    \binoppenalty=10000 %
    \ensuremath{%
      #2%
    }%
    ~\@eqnnum
  \endgroup
}
\makeatother

\usetikzlibrary{shadows}
\definecolor{myyellow}{RGB}{242,226,149}

\NewDocumentCommand\MyStickyNote{O{6cm}mO{6cm}}{%
\begin{tikzpicture}
\node[
drop shadow={
  shadow xshift=2pt,
  shadow yshift=-4pt
},
inner xsep=7pt,
fill=myyellow,
xslant=-0.1,
yslant=0.1,
inner ysep=10pt
] {\parbox[t][#1][c]{#3}{#2}};
\end{tikzpicture}%
}

\makeatletter

\newcommand{\justified}{%
  \rightskip\z@skip%
  \leftskip\z@skip}

\makeatother

\makeatletter \renewcommand{\@dotsep}{2} \makeatother

\makeatletter

\newcommand*\l@chapterinfo{\@nodottedtocline{0}{0.0em}{1.5em}}
\newcommand*\l@sectioninfo{\@nodottedtocline{1}{1.5em}{2.3em}}
\newcommand*\l@subsectioninfo{\@nodottedtocline{2}{3.8em}{3.2em}}
\newcommand*\l@subsubsectioninfo{\@nodottedtocline{3}{7.0em}{4.1em}}
\newcommand*\l@paragraphinfo{\@nodottedtocline{4}{10em}{5em}}
\newcommand*\l@subparagraphinfo{\@nodottedtocline{5}{12em}{6em}}

\def\@nodottedtocline#1#2#3#4#5{%
  \ifnum #1>\c@tocdepth \else
    \vskip \z@ \@plus.2\p@
    {\leftskip #2\relax \rightskip \@tocrmarg \parfillskip -\rightskip
     \parindent #2\relax\@afterindenttrue
     \interlinepenalty\@M
     \leavevmode
     \@tempdima #3\relax
     \advance\leftskip \@tempdima \null\nobreak\hskip -\leftskip
     {#4}\nobreak
     \leaders\hbox{$\m@th
        \mkern \@dotsep mu\hbox{\,}\mkern \@dotsep
        mu$}\hfill
     \nobreak
     \hb@xt@\@pnumwidth{\hfil\normalfont \normalcolor }%
     \par}%
  \fi}

\makeatother

\def\chapterinfo#1{%
    \addcontentsline{toc}{chapterinfo}{%
    \noexpand\numberline{}\parbox[t]{18.00cm}{\justified \fontfamily{lmtt}\selectfont #1}}%
}
\def\sectioninfo#1{%
    \addcontentsline{toc}{sectioninfo}{%
    \noexpand\numberline{}\parbox[t]{16.65cm}{\fontfamily{lmtt}\selectfont #1}}%
}
\def\subsectioninfo#1{%
    \addcontentsline{toc}{subsectioninfo}{%
    \noexpand\numberline{}\parbox[t]{15.30cm}{\fontfamily{lmtt}\selectfont #1}}%
}
\def\subsubsectioninfo#1{%
    \addcontentsline{toc}{subsubsectioninfo}{%
    \noexpand\numberline{}\parbox[t]{13.80cm}{\fontfamily{lmtt}\selectfont #1}}%
}
\def\paragraphinfo#1{%
    \addcontentsline{toc}{parapgraphinfo}{%
    \noexpand\numberline{}\parbox[t]{18.00cm}{\fontfamily{lmtt}\selectfont #1}}%
}
\def\subparagraphinfo#1{%
    \addcontentsline{toc}{subparagraphinfo}{%
    \noexpand\numberline{}\parbox[t]{18.00cm}{\fontfamily{lmtt}\selectfont #1}}%
}

\fancypagestyle{plain}{
    \fancyhf{}
    \fancyfoot[RO,LE]{\thepage}
    \fancyfoot[CO,CE]{}
    \fancyfoot[LO,RE]{}
    
}

\definecolor{gray_title}{gray}{0.5}
\definecolor{gray_text}{gray}{0.75}
\definecolor{gray_line}{gray}{0.90}

\usepackage{indentfirst}

\newdimen\parindentminipage             
\newdimen\parskipminipage 
\newcommand{\vecteur}[1]{\boldsymbol{\mathrm{#1}}}                                      
\newcommand{\diff}[1]{\mathrm{d}#1}                                                     
\newcommand{\partiel}[2]{\frac{\mathrm{\partial}#1}{\mathrm{\partial}#2}}               
\newcommand{\partielsnddouble}[2]{\frac{\mathrm{\partial}^2#1}{\mathrm{\partial}#2^2}}	
\newcommand{\e}[1]{\, \mathrm{e}^{#1}}                                                  

\newcommand{\units}[2]{\ifthenelse{\equal{#2}{}}{\mathrm{#1}}{\mathrm{#1}^{#2}}}

\newcommand{\Smilei}{{\sc Smilei}\xspace}

\makeatletter
 
\newskip\@bigflushglue \@bigflushglue = -100pt plus 1fil

\def\bigcentering{\let\\\@centercr\rightskip\@bigflushglue%
\leftskip\@bigflushglue
\parindent\z@\parfillskip\z@skip}

\newcommand{\documentTitleEn}{
  \begin{center}
    \Large Perfectly Matched Layer implementation for E-H fields and Complex Wave Envelope propagation in the \Smilei PIC code
  \end{center}
}

\newcommand{\Authors}{
  \begin{center}
    Guillaume Bouchard$^{1,2}$, Arnaud Beck$^3$, Francesco Massimo$^4$, Arnd Specka$^3$
  \end{center}
}

\newcommand{\Affiliations}{
  \begin{center}
      $^1$ CEA, DAM, DIF, F-91297 Arpajon, France\\
      $^2$ Université Paris–Saclay, CEA, LMCE, 91680 Bruyères-le-Châtel, France\\
      $^3$ Laboratoire Leprince-Ringuet – École polytechnique, CNRS-IN2P3, Palaiseau 91128, France\\
      $^4$ Laboratoire de Physique des Gaz et des Plasmas - Université Paris-Saclay, CNRS, Orsay 91405, France
  \end{center}
}

\newcommand*\styleSectionA{%
  \titleformat*{\section}{\centering\bf}%
}
\newcommand*\styleSectionB{%
  \titleformat*{\section}{\large\bf}%
}

\begin{document}

\pagestyle{plain}

\documentTitleEn
\Authors
\Affiliations

\styleSectionA
\section*{Abstract}

\noindent
The design of absorbing boundary conditions (ABC) in a numerical simulation is a challenging task.
In the best cases, spurious reflections remain for some angles of incidence or at certain wave lengths.
In the worst, ABC are not even possible for the set of equations and/or numerical schemes used in the simulation and reflections can not be avoided at all.
Perflectly Matched Layer (PML) are layers of absorbing medium which can be added at the simulation edges in order to significantly damp both outgoing and reflected waves, thus effectively playing the role of an ABC.
They are able to absorb waves and prevent reflections for all angles and frequencies at a modest computational cost.
It increases the simulation accuracy and negates the need of oversizing the simulation usually imposed by ABC and leading to a waste of computational resources and power.
PML for finite-difference time-domain (FDTD) schemes in Particle-In-cell (PIC) codes are presented for both Maxwell's equations and, for the first time, the envelope wave equation.
Being of the second order, the latter requires significant evolutions with respect to the former.
It applies in particular to simulations of lasers propagating in plasmas using the reduced Complex Envelope model.
The implementation is done in the open source code \Smilei for both Cartesian and azimuthal modes (AM) decomposition geometries.

\styleSectionB
\section*{Introduction}

In finite-difference time-domain (FDTD) simulations \cite{Yee1966}, boundary conditions are necessary where the stencil used to discretize the local operator would need to access data outside of the simulation domain.
The missing data is given by the boundary conditions in a way that models the expected effects of the exterior onto the simulation domain.

In the case of the so called ``open'' boundary conditions, one expects outgoing waves to cross the transition with the exterior and exit the simulation domain unperturbed and without any reflection back towards the interior.
For Maxwell's equations, several different absorbing boundary conditions (ABC), which implement this behavior, exist.
Examples of this kind of boundary conditions are those known as Silver-Müller \cite{He1999} \cite{baruc1993-1} and Buneman \cite{buneman1993tristan} boundary conditions.
These methods typically evaluate the magnetic field that should be imposed at the boundary in order for an incoming wave to cancel out the outgoing one.
This field is derived from Maxwell's equations and expressed as a function of the local electromagnetic field inside the domain.
Unfortunately, assumptions are made to obtain this function and the resulting boundary is never perfectly absorbing.
Wave reflections remain and  depend on the incidence angle and frequency of the wave.

The Perflectly Matched Layer (PML) is an absorbing media placed at the border of the simulation domain in which outgoing waves are dramatically damped before reaching the actual boundary. 
This damping is tuned to be such that the role of the boundary condition, placed after the PML, becomes almost negligible.
Although the PML is technically not a boundary condition, it is used as such.
It has several advantages over ABC. It is independent of frequency, angle of incidence, polarization and shows much lower reflection errors if the PML parameters are properly tuned.

Berenger's original method for PML is known as the "split field" formulation \cite{BirdsallLangdon2004},\cite{Berenger1994}. In this formulation, the set of equations governing the medium differs from Maxwell's equations. The method is therefore usually referred to as non-Maxwellian and its implementation requires the introduction of new fields. 
Gedney's "unsplit field" formulation or "uniaxial PML", on the opposite, solves the usual Maxwell's equations but in a medium having ad-hoc absorbing properties \cite{Gedney1996-1},\cite{Gedney1996-2}. 
These PML can also be formulated as Maxwell's equations solutions in a space where coordinates are stretched in the complex plane \cite{Chew1994},\cite{Chew1996-1},\cite{Chew1996-2},\cite{Chew1998}.

Between 2000 and 2010, improvements are proposed to increase the accuracy and the stability of PML\cite{kuzuoglu1996,meza2008,meza2010stability}. 
These include the complex frequency shift which is used later in the present work.
In the same period, various implementations are suggested \cite{Roden1997,Roden2010, giannopoulos2008ripml,drossaert2007ripml, ramadan2003-2,Gedney2010} and PML for arbitrary geometries are developed \cite{Teixeira1997-1,Teixeira1997-2,Teixeira1998,Teixeira2000}.
PML is also written for the wave equation \cite{ramadan2003-1} and eventually with the complex representation of fields in the form $F(x,t)=\tilde{F}(x,t)\exp{(-i\omega_0t)}$ \cite{ramadan2007-1,ramadan2007-2,ramadan2008}.
In 2010, implementions of PML, complex frequency shift PML and PML for the second order wave propagation equation appear.
These methods are all written with fields represented in the real space\cite{ma2014-1,ma2014-2,gao2015}.

To the authors' knowledge, no paper reports a general derivation of PML for Maxwell's equations in the time domain for both Cartesian and cylindrical geometry. This paper reports such a derivation and use it to derive PML in the time domain for the solution of Maxwell's equations with the azimuthal modes decomposition \cite{Lifschitz2009}. Additionally, we build on this formulation to derive the PML for the envelope equation of \cite{Massimo2019, Massimo2019cylindrical} in both Cartesian and cylindrical geometries, which was never done in the previous literature.
The paper focuses on how to implement such a formalism in a fully electromagnetic Particle-In-Cell (PIC) code \cite{BirdsallLangdon2004}.
It is a numerical method particularly sensitive to boundary conditions because of the strong coupling between the fields and the macro-particles dynamics and for which no implementation of PML for the envelope wave equation has ever been proposed. 
The first section presents an FDTD derivation of uniaxial PML for standard electromagnetic fields described by Maxwell's equations in the complex coordinate stretching formalism. Section \ref{sec_pml_implementation} then presents how this is implemented in \Smilei \cite{Derouillat2018,Beck2019} in two different geometries and describes benchmarks of these implementations in cases pertaining to high-harmonic generation and laser wakefield acceleration. Finally, Section \ref{sec_wave_propagation} presents the extension of the unsplit PML FDTD formulation to the second-order wave equation for the complex envelope model. 

\section{Uniaxial PML and Complex Coordinate Stretching for Maxwell's equations}\label{sec_pml_derivation}
\subsection{Damping waves in lossy media}\label{sec:dielectric}
This section introduces the general form of Maxwell's equations in an anisotropic and lossy dielectric medium.
It is assumed that any field $\vecteur{F}$ is the sum of monochromatic plane waves $\vecteur{P_j}$ of the form
\begin{equation}\label{eq:plane_wave}
\vecteur{F}(\vecteur{r},t) = \sum_{j}Re\{\tilde{\vecteur{P}}_j \e{i(\omega_j t - \vecteur{k}_j \cdot \vecteur{r})}\},
\end{equation}
where $\tilde{\vecteur{P}}_j$ is the complex amplitude of $\vecteur{P_j}$, a complex vector independent of time.
By linearity of Maxwell's equations, the following derivation is limited to monochromatic plane waves but can be applied to all electromagnetic propagating quantities. 

The Maxwell-Faraday and Maxwell-Ampère equations in an arbitrary medium can be written:
\begin{alignat}{3}\label{eq:Maxwell_vacuum}
& \vecteur{\nabla}\times\vecteur{H} \quad && = \vecteur{J} + \partial_t\vecteur{D} \quad && = [\sigma]\vecteur{E} + \varepsilon_0[\varepsilon_r]\partial_t\vecteur{E},\nonumber\\
& \vecteur{\nabla}\times\vecteur{E} \quad && = -\partial_t\vecteur{B}     \quad && = -\mu_0[\mu_r]\partial_t\vecteur{H},
\end{alignat}
where $\mathbf{E}$ and $\mathbf{H}$ are respectively the electric and magnetic field, $\mathbf{D}$ and $\mathbf{B}$ are respectively the electric displacement and the magnetizing field and $\mathbf{J}$ is the current density.
In Eq. \ref{eq:Maxwell_vacuum}, the following constitutive relations for an anisotropic lossy dielectric medium have been used:
\begin{eqnarray}\label{eq:constitutive}
\vecteur{J} &=& [\sigma]\vecteur{E}, \nonumber \\
\vecteur{D} &=& \varepsilon_0[\varepsilon_r] \vecteur{E},\nonumber\\
\vecteur{B} &=& \mu_0 [\mu_r] \vecteur{H},
\end{eqnarray}
where $\epsilon_0$ and $\mu_0$ are respectively the vacuum permittivity and permeability. $[\sigma]$, $[\varepsilon_r]$ and $[\mu_r]$ are respectively the matrices for the conductivity, relative permittivity and relative permeability of the medium. They are of the form
\begin{equation}
[\sigma] = \begin{pmatrix}
\sigma_{a}& 0 & 0\\
 0 &\sigma_{b}& 0\\
 0 & 0 & \sigma_{c}
\end{pmatrix}
,\qquad
[\varepsilon_r] = \begin{pmatrix}
\varepsilon_{a}&0&0\\
0&\varepsilon_{b}&0\\
0&0&\varepsilon_{c}
\end{pmatrix}
,\qquad
[\mu_r] = \begin{pmatrix}
\mu_{a}&0&0\\
0&\mu_{b}&0\\
0&0&\mu_{c}
\end{pmatrix}.
\end{equation}
If the medium is vacuum, then $[\sigma]=[0]$, $[\varepsilon_r]=[\mu_r]=I_3$, where $I_3$ is the 3$\times$3 identity matrix. From Eq. \ref{eq:plane_wave} and \ref{eq:Maxwell_vacuum} we can write
\begin{eqnarray}
\vecteur{\nabla}\times\vecteur{H}&=&i\omega\epsilon_0[\hat{\varepsilon}]\vecteur{E},\nonumber\\
\vecteur{\nabla}\times\vecteur{E}&=&-i\omega\mu_0[\mu_r]\vecteur{H},\label{eq:Maxwell-fourier}
\end{eqnarray}
with $[\hat{\varepsilon}]$ the complex relative permittivity of the medium defined by
\begin{equation}
[\hat{\varepsilon}]=[\sigma]/i\omega\varepsilon_0 + [\varepsilon_r].\label{eq:eps_complex}
\end{equation}

The propagation along z of a linearly y-polarized wave in such a lossy dielectric medium is then given by:
\begin{equation}
\label{eq:decaying_wave}
E_y = E_{0} \e{i(\omega\varepsilon_b t - k_z z)}\e{\sigma_bt/\varepsilon_{0}},
\end{equation}
which is a decaying solution in time for all $\omega$ if $\sigma_b<0$.
This is the fundamental damping property of the PML which therefore will be built around this concept of lossy dielectric medium with a negative conductivity. 

\subsection{Canceling reflections at the interface}
\begin{figure}
  \includegraphics[scale=1.2]{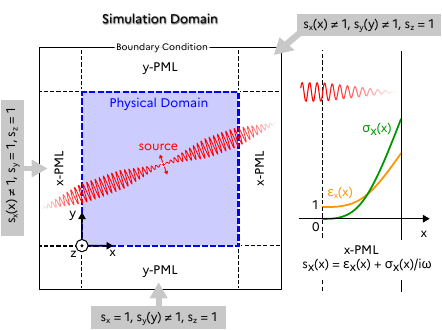}
  \caption{Heuristic representation of PML.
  The simulation domain is composed of the physical domain  and its surrounding PML.
  The so called x and y-PML flank the physical domain respectively along the $x$ and $y$ axes. 
  Electromagnetic waves (in red) are produced by a source placed in the physical domain and propagate until they hit the PML.
  The waves cross the interface and enter the PML without reflection and then are progressively damped before they hit the boundary of the PML.
  A section of the rightmost x-PML along the $x$ axis is detailed on the right panel which represents typical value of $\varepsilon_x(x)$ and $\sigma_x(x)$ as explained in \ref{sec:susceptibility_functions}.}
  \label{fig:pml-scheme}
\end{figure}

The previous section defines how a wave is damped when propagating inside a lossy medium.
In this section, we want to tune the properties of this medium in order to prevent reflections at its interface with the simulation interior as sketched on Fig. \ref{fig:pml-scheme}.
This will conclude the definition of a PML: a lossy dielectric medium into which a wave propagating in the simulation domain can enter without reflection.
If it does, it will later be progressively damped as shown in the previous section and will not perturb the simulation domain.

The first necessary condition for a wave to remain unperturbed at the interface between the simulation interior and the PML is that the two media have the same impedance \cite{Sacks1995}.
The simulation interior area adjacent to the PML is assumed to have the same impedance as vacuum so this condition becomes
\begin{equation}
\frac{\varepsilon_0}{\mu_0}I_3=\frac{\varepsilon_0}{\mu_0}[\hat{\varepsilon]}[\mu_r]^{-1},
\end{equation}
 which imposes $[\mu_r]=[\hat{\varepsilon}]$.
Note that by doing so and following equation \ref{eq:eps_complex}, the permeability of the medium ceases to be real valued.
It follows that we can simplify notations by defining the susceptibility matrix 
\begin{equation}
[s]=[\mu_r]=[\hat{\varepsilon}].
\end{equation}
Reflections at the interface are prevented for all angles and frequencies if the susceptibility $[s]_z$ for a wave propagating in the $z$ direction has the form \cite{Sacks1995},\cite{Gedney1996-1},\cite{Gedney1996-2}

\begin{equation}
\label{eq:complex_permittivity_alongz}
[s]_z=\begin{pmatrix}
s_z&0&0\\
0&s_z&0\\
0&0&s_z^{-1}
\end{pmatrix}.
\end{equation}
This susceptibility is typical of uniaxial crystals \cite{Saleh1991} hence the name of uniaxial PML for the medium inspired from this property.
Following  Eq. \ref{eq:eps_complex}, the whole susceptibility matrix is then defined by only two parameters $\sigma_z$ and $\varepsilon_z$ such as
\begin{equation}
s_z=\frac{\sigma_z}{i\omega\epsilon_0}+\varepsilon_z.
\label{eq:susceptibility}
\end{equation}
For waves propagating along $x$ or $y$, similar conditions are found:
\begin{equation}
[s]_x=\begin{pmatrix}
s_x^{-1}&0&0\\
0&s_x&0\\
0&0&s_x
\end{pmatrix}
,\quad
[s]_y=\begin{pmatrix}
s_y&0&0\\
0&s_y^{-1}&0\\
0&0&s_y
\end{pmatrix}.
\end{equation}
%
%
%

In the general case, several uniaxial media can overlap in the domain corners and the general form of the susceptibility becomes the product of all matrices
\begin{equation}
[s]=\begin{pmatrix}
\dfrac{s_y s_z}{s_x}&0&0\\
0&\dfrac{s_x s_z}{s_y}&0\\
0&0&\dfrac{s_x s_y}{s_z}
\end{pmatrix},\label{eq:pml-medium-matrix}
\end{equation}
where $s_x$,$s_y$,$s_z$ differ from 1 only in areas respectively belonging to an x, y or z-PML as represented on figure \ref{fig:pml-scheme}.
Maxwell's equations in a medium with this susceptibility are given by \\
\parbox{0.5\textwidth}{
\begin{align}
\partiel{E_z}{y}-\partiel{E_y}{z}&=-i\omega \mu_{0} \dfrac{s_ys_z}{s_x} H_x, \nonumber \\[12pt]
\partiel{E_x}{z}-\partiel{E_z}{x}&=-i\omega \mu_{0} \dfrac{s_xs_z}{s_y} H_y, \nonumber \\[12pt]
\partiel{E_y}{x}-\partiel{E_x}{y}&=-i\omega \mu_{0} \dfrac{s_xs_y}{s_z} H_z, \label{eq:maxwell-faraday-uniaxial-pml-medium}
\end{align}
}
\parbox{0.5\textwidth}{
\begin{align}
\partiel{H_z}{y}-\partiel{H_y}{z}&=i\omega \varepsilon_{0} \dfrac{s_ys_z}{s_x}E_x, \nonumber \\[12pt]
\partiel{H_x}{z}-\partiel{H_z}{x}&=i\omega \varepsilon_{0} \dfrac{s_xs_z}{s_y} E_y, \nonumber\\[12pt]
\partiel{H_y}{x}-\partiel{H_x}{y}&=i\omega \varepsilon_{0} \dfrac{s_xs_y}{s_z} E_z. \label{eq:maxwell-ampere-uniaxial-pml-medium}
\end{align}
}

\subsection{Complex coordinate stretching}\label{sec:stretched_coordinates}

In this section we show how the wave propagation in a PML medium is equivalent to the propagation in vacuum in a complex stretched coordinate system, following \cite{Chew1994},\cite{Chew1996-1},\cite{Chew1996-2}.
Let's consider the following change of coordinates
\begin{align}
\tilde{x}&=s_x x\nonumber,\\
\tilde{y}&=s_y y\nonumber,\\
\tilde{z}&=s_z z.
\label{eq:variable_transform}
\end{align}
For the moment, the susceptibility of the medium is assumed to be homogeneous.
Let's consider the following component of Eq. \ref{eq:maxwell-ampere-uniaxial-pml-medium}:
\begin{equation}
\partiel{H_z}{y}-\partiel{H_y}{z}=i\omega \varepsilon_{0} \dfrac{s_ys_z}{s_x}E_x.
\end{equation}
One can rewrite this equation as
\begin{equation}
\frac{1}{s_y}\partiel{(H_z/s_z)}{y}-\frac{1}{s_z}\partiel{(H_y/s_y)}{z}=i\omega \varepsilon_{0} \frac{E_x}{s_x}.
\end{equation}
According to Eq. \ref{eq:variable_transform}, the derivatives can be expressed as derivatives in the new coordinate system
\begin{equation}
\partiel{(H_z/s_z)}{\tilde{y}}-\partiel{(H_y/s_y)}{\tilde{z}}=i\omega \varepsilon_{0} \frac{E_x}{s_x}.
\end{equation}
We now define the following transformed fields
\begin{equation}
\tilde{E_{\zeta}}=E_{\zeta}/s_{\zeta}, \quad
\tilde{H_{\zeta}}=H_{\zeta}/s_{\zeta},
\end{equation}
with $\zeta=x,y,z$. We note that these transformed fields obey standard Maxwell's equations in vacuum in the stretched coordinate system
\begin{equation}
\partiel{\tilde{H_z}}{\tilde{y}}-\partiel{\tilde{H_y}}{\tilde{z}}=i\omega \varepsilon_{0} \tilde{E_x}.
\end{equation}
More generally, we can drop the homogeneous assumption and consider the susceptibility to be a function of space. In that case, $s_\zeta$ becomes a function of coordinate $\zeta$ and the uniaxial PML properties are retrieved with the following stretching
\begin{align}
\tilde{x}(x)&=\int_0^xs_x(x')dx'=x_0+\int_{x_0}^xs_x(x')dx',\\
\tilde{y}(y)&=\int_0^ys_y(y')dy'=y_0+\int_{y_0}^ys_y(y')dy',\\
\tilde{z}(z)&=\int_0^zs_z(z')dz'=z_0+\int_{z_0}^zs_z(z')dz',
\end{align}

where the PML is supposed to start respectively at $x_0$, $y_0$ and $z_0$ so the susceptibility $s_\zeta$ is taken equal to 1 before these points along $x$, $y$ and $z$.
The stretching is only effective inside the PML and leaves the physical domain, where $[s]=I_3$, unmodified.

The wave propagation in a medium with the complex susceptibility of an uniaxial PML defined by Eq \ref{eq:susceptibility} and \ref{eq:pml-medium-matrix} can therefore be seen as the propagation of transformed fields in vacuum in a stretched coordinate system.
This second approach of the PML is convenient because it can be applied to any coordinate system and in particular to the cylindrical coordinate system as described in the next section.
%
%
%

\subsection{Extension of PML for cylindrical coordinates}

This section extends the form of the complex susceptibility that needs to be used for PML to cylindrical coordinates following \cite{Teixeira1997-1},\cite{Teixeira1997-2},\cite{Teixeira1998}.
Using a stretching similar to section \ref{sec:stretched_coordinates}, Maxwell's equations of transformed fields in the stretched cylindrical coordinate system are\\
\parbox{0.5\textwidth}{
\begin{align}
\dfrac{1}{\tilde{r}}\partiel{\tilde{E}_x}{\theta}-\partiel{\tilde{E}_\theta}{\tilde{x}}&=-i\omega \mu_{0} \tilde{H}_r, \nonumber \\[12pt]
\partiel{\tilde{E}_r}{\tilde{x}}-\partiel{\tilde{E}_x}{\tilde{r}}&=-i\omega \mu_{0} \tilde{H}_\theta,\nonumber\\[12pt]
\dfrac{1}{\tilde{r}}\partiel{(\tilde{r} \tilde{E}_\theta)}{\tilde{r}}-\dfrac{1}{\tilde{r}}\partiel{\tilde{E}_r}{\theta}&=-i\omega \mu_{0}  \tilde{H}_x, \label{eq:maxwell-faraday}
\end{align}
}
\parbox{0.5\textwidth}{
\begin{align}
\dfrac{1}{\tilde{r}}\partiel{\tilde{H}_x}{\theta}-\partiel{\tilde{H}_\theta}{\tilde{x}}&=i\omega \varepsilon_{0}  \tilde{E}_r, \nonumber\\[12pt]
\partiel{\tilde{H}_r}{\tilde{x}}-\partiel{\tilde{H}_x}{\tilde{r}}&=i\omega \varepsilon_{0} \tilde{E}_\theta,\nonumber\\[12pt]
\dfrac{1}{\tilde{r}}\partiel{(\tilde{r} \tilde{H}_\theta)}{\tilde{r}}-\dfrac{1}{\tilde{r}}\partiel{\tilde{H}_r}{\theta}&=i\omega \varepsilon_{0} \tilde{E}_x, \label{eq:maxwell-ampere}
\end{align}
}
where
\begin{equation}
\tilde{x} = x_0 + \int_{x_0}^x s_x(x') \,\mathrm{d}x', \qquad
\tilde{r} = r_0 + \int_{r_0}^r s_r(r') \,\mathrm{d}r'. \qquad\label{eq:stretched_xr}
\end{equation}
Note that there is no PML in $\vecteur{e}_\theta$ direction because $\vecteur{e}_\theta$ describes a closed-loop. Electromagnetic waves cannot escape in this direction. Therefore the azimuthal unit vector $\vecteur{e}_\theta$ is not stretched as the radial and longitudinal unit vectors $\vecteur{e}_r$ or $\vecteur{e}_x$. Using again $s_x=d\tilde{x}/dx$ and $s_r=d\tilde{r}/dr$, the differential terms can be expressed as derivatives in the standard coordinate system and the equations take the form \\
%
%
%
\parbox{0.5\textwidth}{
\begin{align}
\dfrac{1}{r}\partiel{(s_x \tilde{E}_x)}{\theta}-\partiel{(\tilde{r}\tilde{E}_\theta/r)}{x}&=-i\omega \mu_{0} \dfrac{\tilde{r}}{r} s_x \tilde{H}_r, \nonumber \\[12pt]
\partiel{(s_r\tilde{E}_r)}{x}-\partiel{(s_x\tilde{E}_x)}{r}&=-i\omega \mu_{0} s_x s_r\tilde{H}_\theta,\nonumber\\[12pt]
\dfrac{1}{r}\partiel{(\tilde{r} \tilde{E}_\theta)}{r}-\dfrac{1}{r}\partiel{(s_r\tilde{E}_r)}{\theta}&=-i\omega \mu_{0}  \dfrac{\tilde{r}}{r}s_r\tilde{H}_x, \label{eq:maxwell-faraday-stretched}
\end{align}
}
\parbox{0.5\textwidth}{
\begin{align}
\dfrac{1}{r}\partiel{(s_x\tilde{H}_x)}{\theta}-\partiel{(\tilde{r}\tilde{H}_\theta/r)}{x}&=-i\omega \mu_{0} \dfrac{\tilde{r}}{r}s_x \tilde{E}_r, \nonumber\\[12pt]
\partiel{(s_r\tilde{H}_r)}{x}-\partiel{(s_x\tilde{H}_x)}{r}&=i\omega \varepsilon_{0} s_x s_r\tilde{E}_\theta,\nonumber\\[12pt]
\dfrac{1}{r}\partiel{(\tilde{r} \tilde{H}_\theta)}{\tilde{r}}-\dfrac{1}{r}\partiel{(s_r\tilde{H}_r)}{\theta}&=i\omega \varepsilon_{0} \dfrac{\tilde{r}}{r} s_r\tilde{E}_x. \label{eq:maxwell-ampere-stretched}
\end{align}
}
Following Eqs. \ref{eq:maxwell-faraday-uniaxial-pml-medium} and \ref{eq:maxwell-ampere-uniaxial-pml-medium}, that were obtained in Cartesian geometry, we are looking for a set of Maxwell's equations in PML for cylindrical geometry of the form

\parbox{0.5\textwidth}{
\begin{align}
\dfrac{1}{r}\partiel{E_x}{\theta}-\partiel{E_\theta}{x}&=-i\omega \mu_{0} \dfrac{s_x s_\theta}{s_r} H_r, \nonumber\\[12pt]
\partiel{E_r}{x}-\partiel{E_x}{r}&=-i\omega \mu_{0} \dfrac{s_x s_r}{s_\theta} H_\theta,\nonumber\\[12pt]
\dfrac{1}{r}\partiel{(r E_\theta)}{r}-\dfrac{1}{r}\partiel{E_r}{\theta}&=-i\omega \mu_{0} \dfrac{s_rs_\theta}{s_x} H_x,
\end{align}
}
\parbox{0.5\textwidth}{
\begin{align}
\dfrac{1}{r}\partiel{H_x}{\theta}-\partiel{H_\theta}{x}&=i\omega \varepsilon_{0} \dfrac{s_x s_\theta}{s_r} E_r,\nonumber\\[12pt]
\partiel{H_r}{x}-\partiel{H_x}{r}&=i\omega \varepsilon_{0} \dfrac{s_x s_r}{s_\theta} E_\theta,\nonumber\\[12pt]
\dfrac{1}{r}\partiel{(r H_\theta)}{r}-\dfrac{1}{r}\partiel{H_r}{\theta}&=i\omega \varepsilon_{0} \dfrac{s_rs_\theta}{s_x} E_x.
\end{align}
}
By identifying this set of equations with Eqs. \ref{eq:maxwell-faraday-stretched} and \ref{eq:maxwell-ampere-stretched}, we find the relations between the transformed and untransformed fields
\begin{align}
E_r = s_r\tilde{E_r},\quad E_\theta = \dfrac{\tilde{r}}{r}\tilde{E}_\theta, \quad E_x = s_x\tilde{E}_x,\nonumber\\
H_r = s_r\tilde{H_r},\quad H_\theta = \dfrac{\tilde{r}}{r}\tilde{H}_\theta, \quad H_x = s_x\tilde{H}_x,
\end{align}
and the the complex susceptibility
\begin{equation}
[s]=\begin{pmatrix}
\dfrac{\tilde{r}}{r}\dfrac{s_x}{s_r}&0&0\\
0&\dfrac{r}{\tilde{r}}s_xs_r&0\\
0&0&\dfrac{\tilde{r}}{r}\dfrac{s_r}{s_x}.
\\
\end{pmatrix}.
\end{equation}
Note that the term $\tilde{r}/r$ can be identified to a formal $s_\theta$ so that $[s]$ eventually takes the same form as in the Cartesian coordinates.

\subsection{Azimuthal modes decomposition}\label{sec:PML_AM}

In \Smilei, the cyclindrical geometry is treated via a decomposition in azimuthal Fourier modes.
Any real scalar field $F\left(x,r,\theta\right)$ (including the components of real vector fields) is decomposed into a basis of azimuthal modes  $\mathcal{F}^m$ with $m$ being the mode number \cite{Lifschitz2009}:
\begin{equation}
F\left(x,r,\theta\right) = \textrm{Re}\left[
    \sum_{m=0}^{+\infty}\mathcal{F}^{m}\left(x,r\right)\exp{\left(-im\theta\right)}
  \right],
\end{equation}
Maxwell's equations for mode $m$ in vacuum is\\
\parbox{0.5\textwidth}{
\begin{align}
-\frac{1}{r}\partial_r(r\mathcal{E}^m_{\theta})-\frac{im}{r}\mathcal{E}^m_r &=\partial_t \mathcal{H}^m_{x},\nonumber\\[12pt]
\frac{im}{r}\mathcal{E}^m_x+\partial_x \mathcal{E}^m_{\theta} &= \partial_t \mathcal{H}^m_r,\nonumber\\[12pt]
-\partial_x \mathcal{E}^m_{r} + \partial_r \mathcal{E}^m_{x} &= \partial_t \mathcal{H}^m_{\theta},
\end{align}
}
\parbox{0.5\textwidth}{
\begin{align}
\frac{1}{r}\partial_r(r\mathcal{H}^m_{\theta})+\frac{im}{r}\mathcal{H}^m_r &= \partial_t \mathcal{E}^m_{x}, \nonumber\\[12pt]
-\frac{im}{r}\mathcal{H}^m_x-\partial_x \mathcal{H}^m_{\theta} &= \partial_t \mathcal{E}^m_r,\nonumber \\[12pt]
\partial_x \mathcal{H}^m_{r} - \partial_r \mathcal{H}^m_{x} &= \partial_t \mathcal{E}^m_{\theta}.
\end{align}
}

It is straightforward to show that this set of equations can be treated as in the previous section and leads to the following set of equations in the PML medium\\
\parbox{0.5\textwidth}{
\begin{align}
\dfrac{1}{r}\partiel{r \mathcal{E}^m_\theta}{r}+\dfrac{im}{r}{\mathcal{E}^m_r}=-i\omega \mu_{0} \dfrac{\tilde{r}}{r}\dfrac{s_r}{s_x} \mathcal{H}^m_x,\nonumber\\[12pt]
-\dfrac{im}{r}\mathcal{E}^m_x-\partiel{\mathcal{E}^m_\theta}{x}=-i\omega \mu_{0} \dfrac{\tilde{r}}{r}\dfrac{s_x}{s_r} \mathcal{H}^m_r,\nonumber\\[12pt]
\partiel{\mathcal{E}^m_r}{x}-\partiel{\mathcal{E}^m_x}{r}=-i\omega \mu_{0} \dfrac{r}{\tilde{r}}s_x s_r \mathcal{H}^m_{\theta}, \label{eq:maxwell-faraday-stretched-mode-decomposition}
\end{align}
}
\parbox{0.5\textwidth}{
\begin{align}
\dfrac{1}{r}\partiel{r \mathcal{H}^m_\theta}{r}+\dfrac{im}{r}\mathcal{H}^m_r=i\omega \varepsilon_{0} \dfrac{\tilde{r}}{r}\dfrac{s_r}{s_x} \mathcal{E}^m_x,\nonumber\\[12pt]
-\dfrac{im}{r}\mathcal{H}^m_x-\partiel{\mathcal{H}^m_\theta}{x}=i\omega \varepsilon_{0} \dfrac{\tilde{r}}{r}\dfrac{s_x}{s_r} \mathcal{E}^m_r,\nonumber\\[12pt]
\partiel{\mathcal{H}^m_r}{x}-\partiel{\mathcal{H}^m_x}{r}=i\omega \varepsilon_{0} \dfrac{r}{\tilde{r}}s_x s_r \mathcal{E}^m_\theta. \label{eq:maxwell-ampere-stretched-mode-decomposition}
\end{align}
}
%
%
%
%
%
%
%

\section{FDTD implementation of uniaxial PML for Maxwell's equations}\label{sec_pml_implementation}

This section describes how the Maxwell's equations in a PML medium, obtained in the previous section, are numerically solved in \Smilei for both Cartesan and cylindrical geometries.
It concludes with two numerical benchmarks illustrating the efficiency of the chosen implementation.

\subsection{Cartesian geometry}

From Eqs. \ref{eq:maxwell-ampere-uniaxial-pml-medium}, Maxwell's equation for the $y$ component of the electric field is
\begin{equation}
\partiel{H_x}{z}-\partiel{H_z}{x}=i\omega\varepsilon_0 \dfrac{s_x s_z}{s_y} E_y \label{eq:maxwell-eq-z-direction}.
\end{equation}
From Eqs. \ref{eq:constitutive} and \ref{eq:pml-medium-matrix}
\begin{equation}\label{eq:constitutive_relation}
D_y = \varepsilon_0\dfrac{s_x s_z}{s_y}E_y.
\end{equation}
And introducing the following normalized electric displacements
\begin{align}
\mathcal{D}_{x}=\frac{D_x}{s_y},
\mathcal{D}_{y}=\frac{D_y}{s_z},
\mathcal{D}_{z}=\frac{D_z}{s_x},
\end{align}
Eq. (\ref{eq:maxwell-eq-z-direction}) can be written as
\begin{equation}
\partiel{H_x}{z}-\partiel{H_z}{x}=i\omega s_z \mathcal{D}_y, 
\end{equation}
yielding, with the definition of $s_z$ in Eq. \ref{eq:susceptibility},
\begin{equation}
\partiel{H_x}{z}-\partiel{H_z}{x} = i\omega \varepsilon_z \mathcal{D}_y + \dfrac{\sigma_z}{\varepsilon_0}\mathcal{D}_y.
\end{equation}
$\mathcal{D}$ is an intermediate quantity stored in memory (in addition to $E$) in the implementation of \Smilei.
Note that this particular choice of normalization is made so that the equations for the $x$ and $z$ components can be easily obtained by simple coordinates permutations.

This equation can be discretized with centered finite differences as in the standard three dimensional Yee finite difference scheme \cite{Yee1966}.
On the right-hand side of the equation, the $i\omega$ term is interpreted as a time differential and the second term must be centered in time leading to
\begin{align}
& \left(
\frac{H_{x_{i,j+1/2,k+1/2}}^n-H_{x_{i,j+1/2,k-1/2}}^n}{\Delta z}
-\frac{H_{z_{i+1/2,j+1/2,k}}^n-H_{z_{i-1/2,j+1/2,k}}^n}{\Delta x}
\right) \nonumber \\
& =
\varepsilon_z \dfrac{\mathcal{D}_{y_{i,j+1/2,k}}^{n+1/2}-\mathcal{D}_{y_{i,j+1/2,k}}^{n-1/2}}{\Delta t} + \dfrac{\sigma_z}{\varepsilon_0}\dfrac{\mathcal{D}_{y_{i,j+1/2,k}}^{n+1/2}+\mathcal{D}_{y_{i,j+1/2,k}}^{n-1/2}}{2},
\end{align}
where the indices $i,j,k$ are the grid point coordinates on the Yee grid.
The previous equation yields an explicit expression for the advancement of $\mathcal{D}_y$ in time
\begin{align}
\mathcal{D}_{y_{i,j+1/2,k}}^{n+1/2} = &
+\frac{2 \varepsilon_0 \varepsilon_z - \Delta t \sigma_z}{2 \varepsilon_0 \varepsilon_z +\Delta t \sigma_z}\mathcal{D}_{y_{i,j+1/2,k}}^{n-1/2}\nonumber \\
&+\frac{2\varepsilon_0 \Delta t}{2 \varepsilon_0 \varepsilon_z + \sigma_z \Delta t}
\left(
\frac{H_{x_{i,j+1/2,k+1/2}}^n-H_{x_{i,j+1/2,k-1/2}}^n}{\Delta z}
-\frac{H_{z_{i+1/2,j+1/2,k}}^n-H_{z_{i-1/2,j+1/2,k}}^n}{\Delta x}
\right).\label{eq:D}
\end{align}
Note that any other form of discretization of the curl of H could be used here if one wants to adapt this implementation to non-standard FDTD scheme.
Using Eq. \ref{eq:constitutive_relation}, the constitutive relation between $\vecteur{E}(t)$and $\vecteur{\mathcal{D}}(t)$,
\begin{equation}
s_y \mathcal{D}_y = \varepsilon_0 s_x E_y,
\end{equation}
which leads to the following equation when introducing the definition of the complex susceptibilities
\begin{equation}
    i\omega \varepsilon_y \mathcal{D}_y + \dfrac{\sigma_y}{\varepsilon_0} \mathcal{D}_y = \varepsilon_0 \left(i\omega\varepsilon_x E_y +\dfrac{\sigma_x}{\varepsilon_0}E_y \right).
\end{equation}
And finally, using a similar discretization gives the explicit expression of the electric field advanced in time.
\begin{equation}\label{eq:evolutionE}
E_y^{n+1/2} = E_y^{n-1/2}\dfrac{\varepsilon_x-\dfrac{\sigma_x\Delta t}{2\varepsilon_0}}{\varepsilon_x+\dfrac{\sigma_x\Delta t}{2\varepsilon_0}} + \dfrac{1}{\left(\varepsilon_x+\dfrac{\sigma_x\Delta t}{2\varepsilon_0}\right)}
\Bigg[\mathcal{D}_y^{n+1/2}\left(\varepsilon_y+\dfrac{\sigma_y\Delta t}{2\varepsilon_0}\right)-\mathcal{D}_y^{n-1/2}\bigg(\varepsilon_y-\dfrac{\sigma_y\Delta t}{2\varepsilon_0}\bigg)\Bigg].
\end{equation}
In order to write the previous equations more compactly, the following coefficients are used\\
\parbox{0.5\textwidth}{
\begin{align}
C_{1,y}(z) &= \frac{2 \varepsilon_0 \varepsilon_z(z) - \sigma_z(z) \Delta t}{2 \varepsilon_0 \varepsilon_z(z) + \sigma_z(z) \Delta t},\nonumber\\[12pt]
C_{2,y}(z) &= \frac{2 \varepsilon_0 \Delta t}{2 \varepsilon_0 \varepsilon_z(z) + \sigma_z(z) \Delta t},\nonumber\\[12pt]
C_{3,y}(x) &= \frac{2 \varepsilon_0 \varepsilon_x(x) - \sigma_x(x) \Delta t}{2 \varepsilon_0 \varepsilon_x(x) + \sigma_x(x) \Delta t},\nonumber
\end{align}
}
\parbox{0.5\textwidth}{
\begin{align}
C_{4,y}(x) &= \frac{1}{2 \varepsilon_0 \varepsilon_x(x) + \sigma_x(x) \Delta t},\nonumber\\[20pt]
C_{5,y}(y) &= 2 \varepsilon_0 \varepsilon_y(y) + \sigma_y(y) \Delta t,\nonumber\\[24pt]
C_{6,y}(y) &= 2 \varepsilon_0 \varepsilon_y(y) - \sigma_y(y) \Delta t\nonumber.
\end{align}
}\\
The $y$ subscripts indicate that these coefficients apply to the $y$ component of the fields in the considered PML.
We recall here that the conductivity and permittivity parameters are functions of the position in the considered PML and so are the coefficients $C_{1-6}$.
Eqs. \ref{eq:D} and \ref{eq:evolutionE} then become
\begin{align}
\mathcal{D}_{y_{i,j+1/2,k}}^{n+1/2} = &C_{1,y}(z)\mathcal{D}_{y_{i,j+1/2,k}}^{n-1/2}\nonumber\\
+ &C_{2,y}(z)
\left(
\frac{H_{x_{i,j+1/2,k+1/2}}^n-H_{x_{i,j+1/2,k-1/2}}^n}{\Delta z}
-\frac{H_{z_{i+1/2,j+1/2,k}}^n-H_{z_{i-1/2,j+1/2,k}}^n}{\Delta x}
\right),\label{eq:Dcompact} \\
E_y^{n+1/2} = &
C_{3,y}(x) E_y^{n-1/2} + C_{4,y}(x)\label{eq:Ecompact}
\Bigg[C_{5,y}(y)\mathcal{D}_y^{n+1/2}-C_{6,y}(y)\mathcal{D}_y^{n-1/2}\Bigg].
\end{align}
In \Smilei, coefficients $C_{1-6}$ are computed and stored once for each component $x,y,z$ and for each PML.
Eqs. \ref{eq:Dcompact} and \ref{eq:Ecompact} are evaluated at each time step in order to advance $E_y$ in time inside the PML domains.
For $B_y$ and $H_y$ the same coefficients are used but they have to be evaluated at different positions on the Yee grid.
In practice, each coefficient is evaluated on both the primal (integer) and dual (half integer) indices which leads to a total of 36 coefficients arrays stored per PML (6 coefficicients for 3 components on 2 meshes).

The derivation described here is done for a $y$ component of the field.
$x$ and $z$ components are obtained by a simple circular permutation over the coordinates.
The value of $\varepsilon$ and $\sigma$ is left to the user but guidlines are given in section \ref{sec:susceptibility_functions}.



\subsection{Cylindrical geometry}

In this geometry dimensions can not be treated indifferently and the PML conditions for one direction cannot be easily retrieved from those of another direction by simple permutations.
Therefore the formulations for all components are given below.
In cylindrical geometry the chosen normalization is
\begin{equation}
\mathcal{D}_r = s_\theta \mathcal{E}_r / s_r , \qquad \mathcal{D}_x = s_\theta \mathcal{E}_x / s_x, \, \text{and} \qquad \mathcal{D}_\theta = s_x \mathcal{E}_\theta / s_\theta.  
\end{equation}

Then Eqs. \ref{eq:maxwell-ampere-stretched-mode-decomposition} becomes
%
%
\begin{align}
\mathcal{D}_r^{n+1/2} &= C_{1,r}(x) \mathcal{D}_r^{n-1/2} + C_{2,r}(x) \bigg[ -\dfrac{im}{r}\mathcal{H}_x-\partiel{\mathcal{H}_\theta}{x} \bigg],\\[12pt]
\mathcal{D}_x^{n+1/2} &= C_{1,x}(r) \mathcal{D}_x^{n-1/2} + C_{2,x}(r) \bigg[ \dfrac{1}{r}\partiel{r \mathcal{H}_\theta}{r}+\dfrac{im}{r}\mathcal{H}_r \bigg], \\[12pt]
\mathcal{D}_\theta^{n+1/2} &= C_{1,\theta}(r) \mathcal{D}_\theta^{n-1/2} + C_{2,\theta}(r) \bigg[ \partiel{\mathcal{H}_r}{x}-\partiel{\mathcal{H}_x}{r} \bigg],
\end{align}
%
%
%
and the electric field is updated following Eqs. \ref{eq:maxwell-faraday-stretched-mode-decomposition}, which become
%

%
%
\begin{align}
\mathcal{E}_r^{n+1/2} = C_{3,r}(r) \mathcal{E}_r^{n-1/2} + C_{4,r}(r) \bigg[ C_{5,r}(r) \mathcal{D}_r^{n+1/2} - C_{6,r}(r) \mathcal{D}_r^{n-1/2} \bigg],\\[12pt]
\mathcal{E}_x^{n+1/2} = C_{3,x}(r) \mathcal{E}_x^{n-1/2} + C_{4,x}(r) \bigg[ C_{5,x}(x) \mathcal{D}_x^{n+1/2} - C_{6,x}(x) \mathcal{D}_x^{n-1/2} \bigg], \\[12pt]
\mathcal{E}_\theta^{n+1/2} = C_{3,\theta}(x) \mathcal{E}_\theta^{n-1/2} + C_{4,\theta}(x) \bigg[ C_{5,\theta}(r) \mathcal{D}_\theta^{n+1/2} - C_{6,\theta}(r) \mathcal{D}_\theta^{n-1/2} \bigg].
\end{align}
The coefficients $C$ are \\
\parbox{0.5\textwidth}{
\begin{align}
C_{1,r}(x) &= \frac{2 \varepsilon_0 \varepsilon_x(x) - \sigma_x(x) \Delta t}{2 \varepsilon_0 \varepsilon_x(x) + \sigma_x(x) \Delta t},\\[12pt]
C_{2,r}(x) &= \frac{2 \varepsilon_0 \Delta t}{2 \varepsilon_0 \varepsilon_x(x) + \sigma_x(x) \Delta t},\\[12pt]
C_{3,r}(r) &= \frac{2 \varepsilon_0 (r_0+I_\varepsilon(r)) - I_\sigma(r) \Delta t}{2 \varepsilon_0 (r_0+I_\varepsilon(r)) + I_\sigma(r) \Delta t},
\end{align}
}
\parbox{0.5\textwidth}{
\begin{align}
C_{4,r}(r) &= \frac{r}{2 \varepsilon_0 (r_0+I_\varepsilon(r)) + I_\sigma(r) \Delta t},\\[20pt]
C_{5,r}(r) &= 2 \varepsilon_0 \varepsilon_r(r) + \sigma_r(r) \Delta t, \\[28pt]
C_{6,r}(r) &= 2 \varepsilon_0 \varepsilon_r(r) - \sigma_r(r) \Delta t, \\[-10pt]\nonumber
\end{align}
}\\
%
%
\parbox{0.5\textwidth}{
\begin{align}
C_{1,x}(r) &= \frac{2 \varepsilon_0 \varepsilon_r(r) - \sigma_r(r) \Delta t}{2 \varepsilon_0 \varepsilon_r(r) + \sigma_r(r) \Delta t},\\[12pt]
C_{2,x}(r) &= \frac{2 \varepsilon_0 \Delta t}{2 \varepsilon_0 \varepsilon_r(r) + \sigma_r(r) \Delta t},\\[12pt]
C_{3,x}(r) &= \frac{2 \varepsilon_0 (r_0+I_{\varepsilon}(r)) - I_\sigma(r) \Delta t}{2 \varepsilon_0 (r_0+I_{\varepsilon}(r)) + I_\sigma(r) \Delta t},
\end{align}
}
\parbox{0.5\textwidth}{
\begin{align}
C_{4,x}(r) &= \frac{r}{2 \varepsilon_0 (r_0+I_{\varepsilon}(r)) + I_\sigma(r) \Delta t},\\[20pt]
C_{5,x}(x) &= 2 \varepsilon_0 \varepsilon_x(x) + \sigma_x(x) \Delta t,  \\[28pt]
C_{6,x}(x) &= 2 \varepsilon_0 \varepsilon_x(x) - \sigma_x(x) \Delta t,  \\[-10pt]\nonumber
\end{align}
}\\
%
%
\parbox{0.5\textwidth}{
\begin{align}
C_{1,\theta}(r) &= \frac{2 \varepsilon_0 \varepsilon_r(r) - \sigma_r(r) \Delta t}{2 \varepsilon_0 \varepsilon_r(r) + \sigma_r(y) \Delta t},\\[12pt]
C_{2,\theta}(r) &= \frac{2 \varepsilon_0 \Delta t}{2 \varepsilon_0 \varepsilon_r(r) + \sigma_r(r) \Delta t},\\[12pt]
C_{3,\theta}(x) &= \frac{2 \varepsilon_0 \varepsilon_x(x) - \sigma_x(x) \Delta t}{2 \varepsilon_0 \varepsilon_x(x) + \sigma_x(x) \Delta t},
\end{align}
}
\parbox{0.5\textwidth}{
\begin{align}
C_{4,\theta}(x) &= \frac{1}{2 \varepsilon_0 \varepsilon_x(x) + \sigma_x(x) \Delta t},\\[12pt]
C_{5,\theta}(r) &= \frac{2 \varepsilon_0 (r_0 + I_\varepsilon(r)) + I_\sigma(r) \Delta t}{r},\\[12pt]
C_{6,\theta}(r) &= \frac{2 \varepsilon_0 (r_0+I_\varepsilon(r)) - I_\sigma(r) \Delta t}{r},
\end{align}
}\\
where $I_{\sigma}(r)=\int_{r_0}^r\frac{\sigma_r(r')}{i\omega\varepsilon_0}dr'$ and $I_{\varepsilon}(r)=\int_{r_0}^r\varepsilon_r(r')$.
These terms appear in the definition of $\tilde{r}$ according to its definition in Eq. \ref{eq:stretched_xr}.
Let's recall here also that $r_0$ is the radial coordinate of the beginning of the radial PML.
Similarly to the Cartesian geometry, each coefficient must be evaluated at the field's location on the grid and the same coefficients are used in order to evaluate $\vecteur{H}$.

\subsection{Conductivity and permittivity functions}\label{sec:susceptibility_functions} 

The derivation of the PML equations does not make any assumption on the shape of the $\sigma$ and $\varepsilon$ functions.
They define the variation of the physical properties of the PML medium, as described in section \ref{sec:dielectric}, and have not been constrained so far.
Nevertheless, in the discrete implementation of the method, the shapes of these functions are important.
Empirical results tend to show that they are better chosen monotonic, varying slowly and being equal to the vacuum property at the boundary with the physical domain in order to have a smooth transition \cite{gedney2022introductionfdtd}. A too high or non continuous absorption can lead to numerical errors because the FDTD scheme becomes unable to evaluate correctly the local derivatives of the fields.

To the knowledge of the authors, there is no demonstrated optimal configuration adapted to all physics cases.
A common recommendation \cite{Gedney1996-1,Gedney1996-2,gedney2022introductionfdtd} is given by the following polynomial functions
\begin{equation}\label{eq:sigma_and_epsilon}
\sigma(x)= \left(\frac{x}{d}\right)^m\sigma_{\rm max},\quad
\varepsilon(x)= 1 +\left(\frac{x}{d}\right)^m(\varepsilon_{\rm max}-1),
\end{equation}
where $x$ is the coordinate along the normal direction of the PML. By convention here the PML starts at $x=0$, is defined for $x>0$ and ends at $x=d$ which is the size of the PML given by $d=N_{\rm PML cells} \times \Delta x$. The parameters $\sigma_{\rm max}$ and $\varepsilon_{\rm max}$ are the maximum conductivity and permittivity reached. Their values as well as the one of the polynomial order $m$ are discussed in the following.

Note that the effective absorption of the PML depends on $\sigma$: the higher the value of $\sigma$, the greater the damping.
But it also depends on the physical width $d$ of the PML because waves are progressively damped during their propagation.
This means that increasing the simulation resolution without increasing the number of PML cells would result in thinner PML and therefore  lower absorption if all other parameters remain unchanged.
Also note that using a too large $\sigma_{\rm max}$ can lead to the discretization errors mentioned previously.
Reversely, if it is too small, the incoming wave is not fully absorbed by the PML.
The theoretical $\sigma_{\rm max}$ required for a given absorption coefficient $R$ for a normally incident wave \cite{gedney2022introductionfdtd} is evaluated as

\begin{equation}\label{eq:sigma_max}
\sigma_{\rm max} = -\frac{(m+1)\ln{R}}{2d}.
\end{equation}

In \Smilei, the default $\sigma$ function is chosen as a polynomial of order $m=2$ and $\sigma_{\rm max}=20$.
These values are obtained by applying equation \ref{eq:sigma_max} to a simulation with 10 PML cells, a resolution of $\Delta x=0.1$ and a target absorption coefficient of $R\approx10^{-6}$ which should be sufficient for most PIC simulations even though these choices remain partly empirical. 

The $\varepsilon$ parameter allows for a better absorption of evanescent waves and waves with a grazing incidence \cite{gedney2022introductionfdtd,petropoulos2000,zhang2010} but is subject to the same kind of limitation due to the introduction of discretization errors if $\varepsilon$ is too high. In \Smilei, the default $\varepsilon$ is empirically chosen to be a polynomial function of order $m=4$ and $\varepsilon_{\rm max}=80$.
Note that the $\sigma$ and $\varepsilon$ profiles can be fully defined by the user.

\subsection{Benchmarks}

This section illustrates the efficiency of the PML implementation in \Smilei for production cases both in Cartesian and cylindrical geometries.

\subsubsection{X-UV emission from laser and overdense plasma foil interaction}

\begin{figure}
  \includegraphics[scale=0.650]{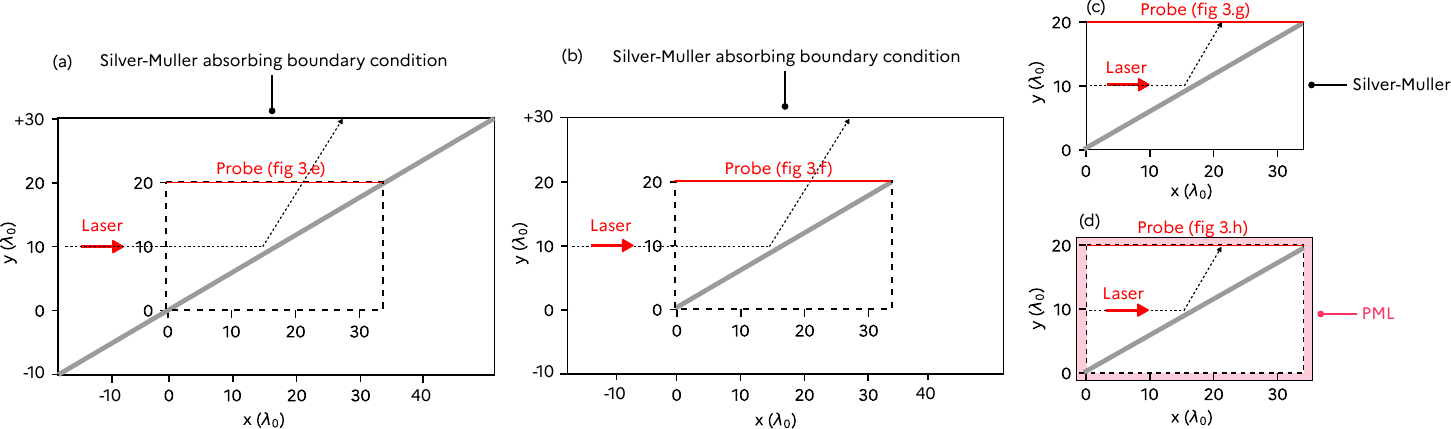}
  \caption{Configurations of the 4 simulations testing the efficiency of the Cartesian PML implementation in \Smilei. The simulations in panels a, b have a larger simulation domain in order to avoid any influence of the boundary on the results on axis and differ only by the size of the plasma slab represented as a thick gray line. The simulations in panels c, d have a smaller simulation domain and are thus exposed to potential reflections on the upper boundary. They differ only by the type of boundary conditions.
  The laser optical trajectory is represented as a dotted line.
  The position of the probe from which $B_z$ is measured and shown on the bottom panels of Fig. \ref{fig:Cartesian_pml} is represented as a red line.
  Panels a,b only show a spatial subset of the simulation domain, equal to the simulation domain shown in panels c,d.} 
  \label{fig:Cartesian_pml-scheme}
\end{figure}

This first benchmark features a typical ultra high intensity (UHI) configuration where an intense laser pulse hits a high density target with an angle \cite{Kim2023, Fedeli2021}.
The laser has a carrier wavelength $\lambda_0=0.8$ $\mu$m and normalized peak electric field $a_0=10$ at waist in vacuum.
The normalized peak electric field is defined as $a_0=max[|\mathbf{E}|]/(m_e\omega_0c/e)$, where $|\mathbf{E}|$ is the absolute value of the laser field at the focal plane, $m_e$ is the electron mass, $\omega_0=2\pi c/\lambda_0$ is the laser carrier angular frequency and $c$ is the speed of light.
The laser pulse waist is $w=3\lambda_0$, and its FWHM duration in intensity is $T_{fwhm} = 25$ fs with Gaussian temporal and transverse profiles.
It propagates along the x direction, impinges on a solid and thin target of thickness $L=0.7\lambda_0$ with an angle of incidence $\theta_{inc}=60^{\circ}$ and is reflected towards the upper boundary of the domain. The solid target is chosen to be aluminium which has an electronic density $n_e = 576\thinspace n_{crit}$ at this wavelength, $n_{crit}=(\varepsilon_0\omega_0^2m_e)/e^2$ being the critical density for a laser of angular frequencey $\omega_0$.

Four Cartesian two dimensional runs are carried out with various global simulation sizes, plasma sizes and  boundary conditions.
For all cases, the simulation time is $T_{sim} = 60 \lambda_0/c$. Spatial and temporal resolutions are $\Delta x = \Delta y = \lambda_0/128$, $c \Delta t = \Delta x / 2$. 1000 macro-particles per cell are used in the plasma slab.

The so called "Reference" simulations (Figs.\ref{fig:Cartesian_pml-scheme} a, b) have physical domain with size of $68\ \lambda_0$ along $x$ and $40\, \lambda_0$ along $y$.
These two cases are large enough so that reflections at the boundaries do not affect the results near the optical axis at the center of the domain. Both are carried out with Silver-Müller \ boundary conditions and without PML. The simulation in Fig.\ref{fig:Cartesian_pml-scheme}a has a plasma slab in the whole extent of the domain whereas the simulation in Fig. Fig.\ref{fig:Cartesian_pml-scheme}b has a reduced plasma slab of the same size as the one in the simulations with the "reduced" physical domain (shown in Figs.\ref{fig:Cartesian_pml-scheme}c, d).
The "reduced" simulations in Figs.\ref{fig:Cartesian_pml-scheme}c,d have a physical domain with smaller size, i.e. $34\ \lambda_0$ along $x$ and $20\ \lambda_0$ along $y$ and the plasma slab extends across the whole simulation domain.
Within such smaller domains, reflections at the $y$ boundary could potentially affect the results near the optical axis.
The simulation in Fig.\ref{fig:Cartesian_pml-scheme}c uses Silver-Müller absorbing boundary conditions \cite{baruc1993-1,Barucq1993-2}.
The simulation in Fig.\ref{fig:Cartesian_pml-scheme}d uses 20 PML cells instead.
$\sigma$ and $\varepsilon$ are polynomials of respective order $m=2$ and $m=4$, $\sigma_{\rm max}=160$ and $\varepsilon_{\rm max}=80$ (see Eq. \ref{eq:sigma_max}).
Fig. \ref{fig:Cartesian_pml-scheme} displays an overview of the 4 simulations.

\begin{figure}
    \includegraphics[width=0.9\textwidth]{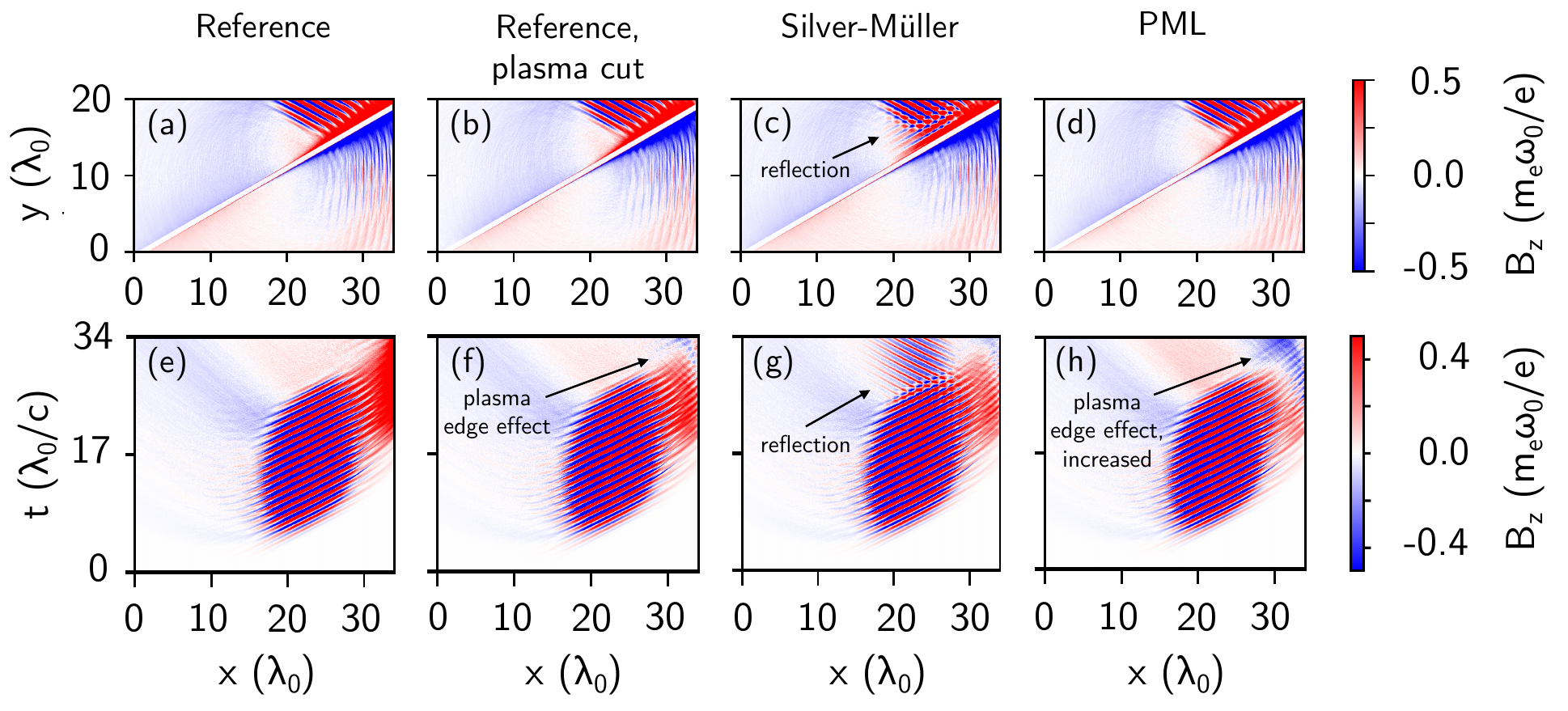}
  \caption{Top panels (a-d) display the laser field $B_z$ for all 4 corresponding simulations as a function of spatial coordinates $x$ and $y$ at time $t=44\lambda_0/c$. Bottom panels (e-h) display the same laser field of the respective (a-d) simulations as a function of $x$ and time $t$ at $y=19.99\lambda_0$ (see probe position on Fig. \ref{fig:Cartesian_pml-scheme}). Panels (a) and (e) display the reference results. Simulation (b) shows that forcing a sharp plasma edge, even in a large domain in which boundary conditions are out of cause, triggers some noise visible on panel (f). Panels (c) and (g) display a large increase in reflection caused by the use of a smaller domain with Silver-Müller absorbing boundary conditions, which are known to be less efficient when outgoing fields hit the border with an angle. Finally, panels (d) and (h) show that the use of PML allows to recover almost perfectly the reference simulation (b).}
  \label{fig:Cartesian_pml}
\end{figure}

In all simulations, the laser, propagating initially along $x$, hits the target at an angle of $\pi/3$ before being reflected towards the $y$ direction.
The laser field $B_z$ for all simulations is shown on Fig. \ref{fig:Cartesian_pml}.
In the Reference cases, the laser exits the central region without perturbing the field on target.
This is not true anymore in the smaller domain simulation with Silver-Müller ABC shown in Fig. \ref{fig:Cartesian_pml-scheme}c. 
In that case, as the laser crosses the $y$ boundary, a part of the laser pulse is reflected back towards the target.
These reflections do not occur when using PML as seen on Fig. \ref{fig:Cartesian_pml-scheme}d.
Note that the Reference simulation in panel \ref{fig:Cartesian_pml-scheme}f displays some noise, with respect to panel \ref{fig:Cartesian_pml-scheme}e,  due to the presence of a sharp plasma edge.
This is independent of the boundary conditions, which are located too far away to be responsible for this noise at such an early time.
Therefore, this reflection at the plasma edge is physically accurate.
Nevertheless, it appears that this noise is slightly enhanced in the case of the smaller domain with PML.
This may be due to the fact that the edge is smoothed in the case of the Reference simulation where the domain is larger and the edge is sharper in the case of the small simulation domain used in the PML case. 
Even though it could be improved by a better choice of $\sigma$ and $\varepsilon$ in this particular case, it is a typical example of the most important shortcoming of the PML method which is that they are derived by assuming an interface with vacuum.
This assumption is not accurate when there is a significant plasma flow close to the boundary \cite{PML_lehe}.
It is therefore good practice to use density profiles within the physical domain that progressively reaches zero close to the PML region whenever possible.

Eventually, a simulation domain smaller by a factor of 4 in total number of cells and the use of PML are able to give results in very good agreement with the reference simulation. 
This illustrates the efficiency of the method and leads to potential significant computational resources savings.


\subsubsection{Cylindrical geometry}\label{sec:benchmark_standard_cylindrical_PML}

\begin{figure}[h!]
    \includegraphics[width=\textwidth]{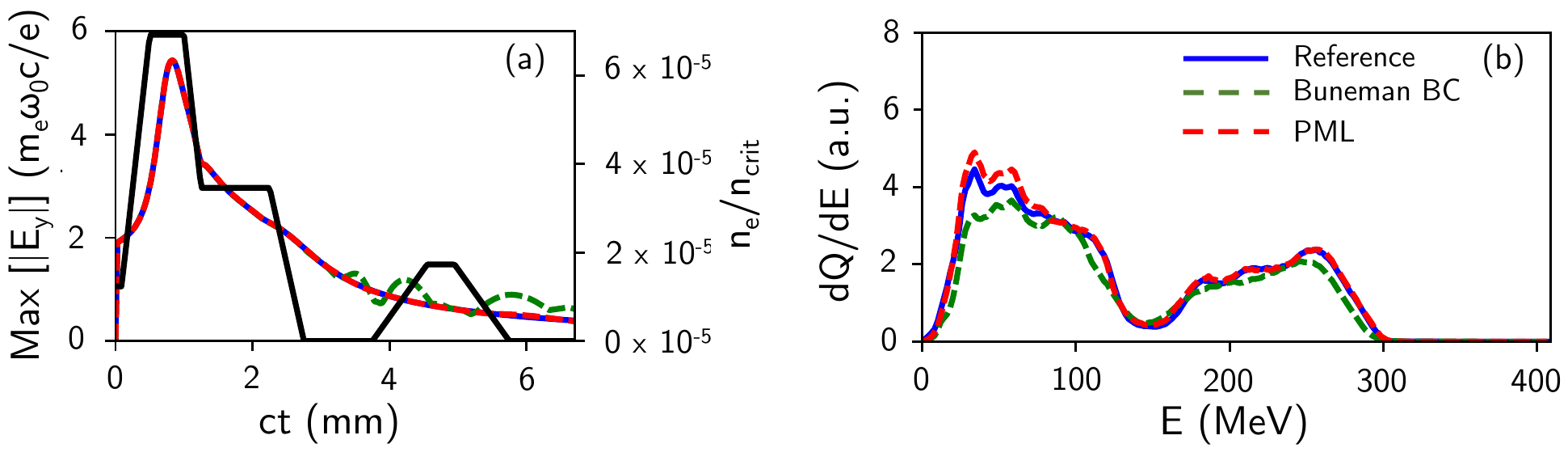}

  \caption{Results of Reference, Buneman BC and PML simulations. Panel a): evolution of the peak laser field $max(|\mathbf{E}|)$. The black line shows the plasma density profile.
  Panel b): electron spectra obtained in the three simulations after 6 mm of propagation. Only the electrons obtained from ionization are shown.
  }\label{fig:cylindrical_standard_a0_and_spectrum}
\end{figure}

A case of laser wakefield acceleration (LWFA) \cite{Esarey2009} is described in this section as a benchmark for the PML implementation in cylindrical geometry with azimuthal Fourier modes decomposition described in section \ref{sec:PML_AM}.
An intense laser pulse is focused into a transversely uniform plasma target of undercritical density (i.e. with density smaller than $n_{crit}$) and composed of fully ionized hydrogen and a 2\% dopant made of $N^{5+}$ ions.
The longitudinal density profile is tailored with two successive plasmas separated by vacuum as shown on Fig.\ref{fig:cylindrical_standard_a0_and_spectrum}a.
The first plasma is used to inject electrons stripped from the nitrogen ions via ionization process \cite{Chen2012} into the plasma wave generated by the laser pulse during its propagation.
Later, the beam formed by these electrons is accelerated in a region with constant density and exits the plasma after the first density downramp.
After exiting the plasma, the accelerated electron beam propagates in vacuum and eventually enters the second plasma which has lower density and acts as a plasma lens \cite{Lehe2014}.

The laser pulse, with carrier wavelength $\lambda_0=0.8$ $\mu$m and linearly polarized along the $y$ direction, has a Gaussian profile in the longitudinal and transverse directions, with FWHM duration in intensity $30$ fs, waist $w_0=15$ $\mu$m and normalized peak electric field $a_0=2.2$ at waist in vacuum.
The laser is injected from the left border of the simulation window, with focal plane position in vacuum located at the end of the upramp.
The first plasma target has a first plateau density of $n_1=6\cdot10^{18}$ cm$^{-3}$ and a second plateau density of $n_2=3\cdot10^{18}$ cm$^{-3}$.
The length of its upramp, of its first plateau region, of its first downramp, of its second plateau region and of its second downramp are respectively 500 $\mu$m, 500 $\mu$m , 250 $\mu$m, 1000 $\mu$m, 500 $\mu$m. After a vacuum region of length 1 mm, the upramp of the second plasma starts.
The second plasma target has a density upramp of length 800 $\mu$m, a plateau of density with length 400 $\mu$m and density $n_3=1.5\cdot10^{18}$ cm$^{-3}$, and a downramp of length 800 $\mu$m.

The results of three simulations are compared: one with a transversely large window and ABC of type Buneman \cite{buneman1993tristan} at the radial boundary to avoid the laser reflections during its propagation (referred to as Reference simulation); a simulation with a significantly smaller transverse size using similar ABC boundary conditions (referred to as Buneman BC simulation) and a simulation with the same smaller size but using radial PML boundary conditions (referred to as PML simulation). 
Using a moving window, the simulation domain is constantly shifting forward and follows the laser pulse, therefore the boundary conditions at the longitudinal boundaries have insignificant influence on the results.
This setup therefore focuses on the radial boundaries, specific to this geometry.
The three simulations all use 2 azimuthal modes, with a resolution of $\Delta x=0.25$ $c/\omega_0$ and $\Delta r=1.0$ $c/\omega_0$ in the $x$ and radial $r$ directions respectively. The timestep for the three simulations has been set to $\Delta t = 0.225$ $\omega_0^{-1}$. The longitudinal length of the moving window for the three simulations is $L_x=704$ $c/\omega_0$. The transverse size $L_r$ of the window is 1408 $c/\omega_0$, 512 $c/\omega_0$, 512 $c/\omega_0$ for the reference, Buneman and PML simulation respectively.
To avoid abrupt reflections at the boundaries of the smaller simulation domains as explained in the previous section, the plasma density decreases linearly to zero from $r=416$ $c/\omega_0$ to $r=512$ $c/\omega_0$ in these two cases.
The three simulations use 32 macro-particles per cell ($[1,4,8]$ regularly spaced respectively in the $x$, $r$ directions and along the azimuthal angle $\theta=[0,2\pi]$) to sample the background electrons and 8 macro-particles per cell ($[1,1,8]$) to sample the $N^{5+}$ ions .

\begin{figure}[h!]
    \includegraphics[width=\textwidth]{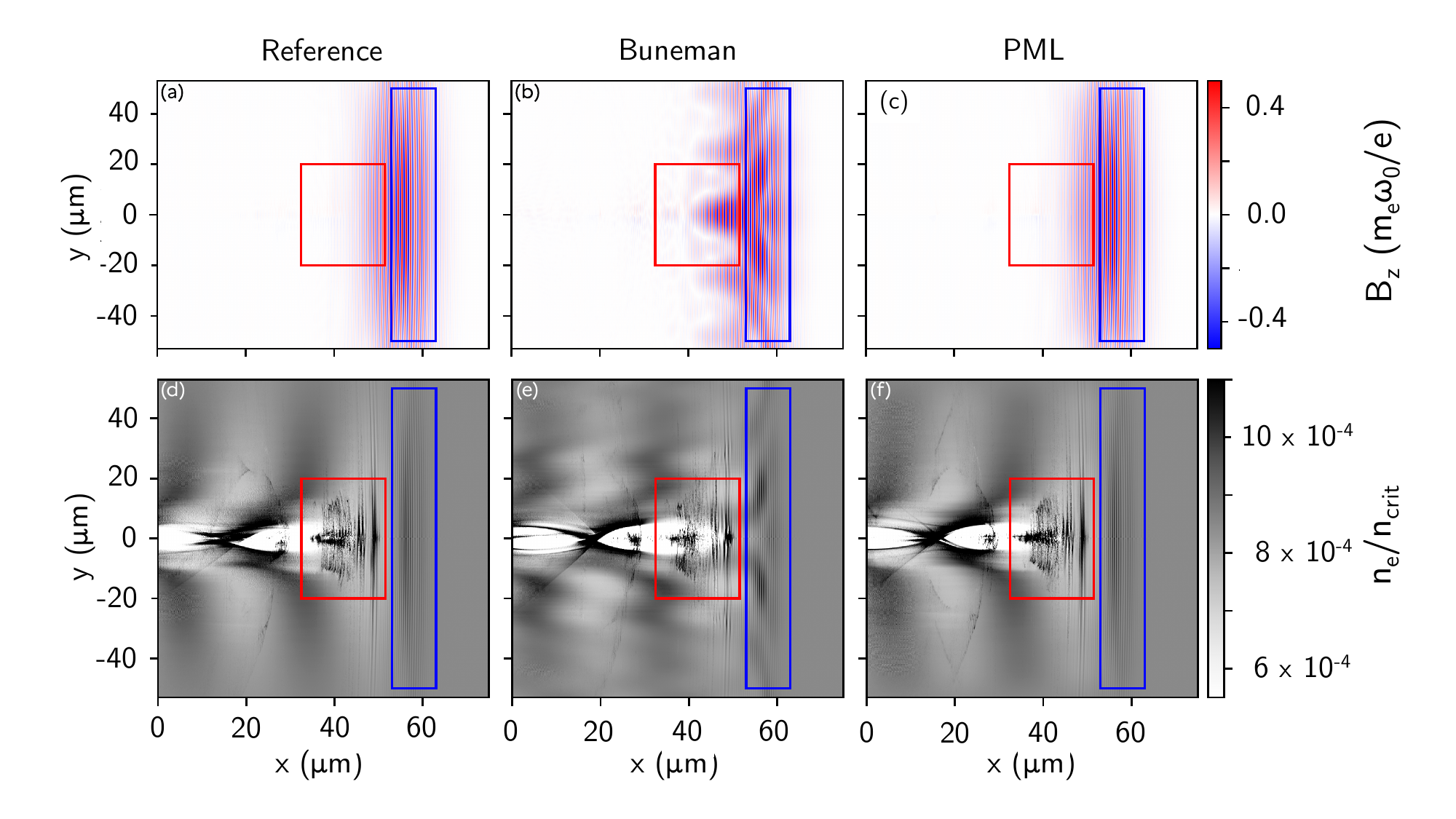}
  \caption{Results of Reference, Buneman BC and PML simulations when the beam reaches the middle of the second plasma after 166 000 iterations, corresponding to 4.75 mm of propagation. Top panels: Bz field on the plane xy; Bottom panels: electron density on the plane xy.
  Panels a),d): Reference simulation; Panels b),e): Buneman simulation; Panels c),f) PML simulation. For readers' convenience, a blue rectangle was added in the same zone where the peak field of the laser pulse lies; similarly, a red rectangle was added to mark the position of the main accelerated electron beam.}\label{fig:Bz_and_density_cylindrical_standard}
\end{figure}

Fig. \ref{fig:Bz_and_density_cylindrical_standard} (top panel) shows the laser $B_z$ field on the $xy$ plane when the beam reaches the middle of the second plasma after 166 000 iterations equivalent to a propagation of 4.75 mm of propagation in the three simulations.
In the Buneman simulation the laser reflections at the border have a significant effect on the laser propagation, and the field distribution is visibly different from the one in the Reference simulation.
Furthermore, lateral lobes where the field is more intense have appeared in the interference pattern. On the other hand, the field in the PML simulations is very similar to the one in the Reference simulation.
The effects of these results on the laser field are shown in the bottom panel of Fig. \ref{fig:Bz_and_density_cylindrical_standard}, where the total electron density is shown.
The laser field lateral lobes found in the Buneman simulation completely alter the density distribution of the wave around the laser wakefield, and the intense field in the electron beam region (marked in a red box) in the Buneman simulation disrupts the dynamics of the beam electrons.

Fig. \ref{fig:cylindrical_standard_a0_and_spectrum} (left panel) shows the evolution of the peak electric field $max(|\mathbf{E}|)$ (which is dominated by the laser field) during its the propagation.
For the reader's convencience, also the unperturbed plasma density profile is shown.

In all three simulations the laser pulse is subject to a phase of relativistic self-focusing \cite{Sun1987} followed by a phase where the laser diffracts.
In the first plasma target, the evolution of the peak electric field is essentially the same in the three simulations. 
The Buneman boundary conditions are sufficient to absorb the laser pulse field at the borders.
However, in the vacuum region between the two plasmas, the first differences appear between the Reference simulation and the Buneman BC simulation.
The latter displays oscillations in the peak electric field because the reflections of the laser at the radial boundary eventually reach the axis and alter the peak field.
The results of the PML simulation on the other hand keep mirroring those of the Reference simulation.
The laser pulse has been absorbed at the transverse boundary and no alterations to the peak field are observed. 

As discussed before, the reflections of the laser pulse alter the distribution of the wakefield, and these differences have significant effects on the electron dynamics in the laser wakefield. Fig. \ref{fig:cylindrical_standard_a0_and_spectrum} (right panel) shows the energy spectrum for the three simulations after 6 mm, in particular for the electrons obtained from ionization.
From this figure it can be inferred that, compared to the results of the Buneman simulation, those of the PML simulation are more similar, almost superposed, to the spectrum obtained in the Reference simulation.

\begin{figure}[hb!]
    \includegraphics[width=0.9\textwidth]{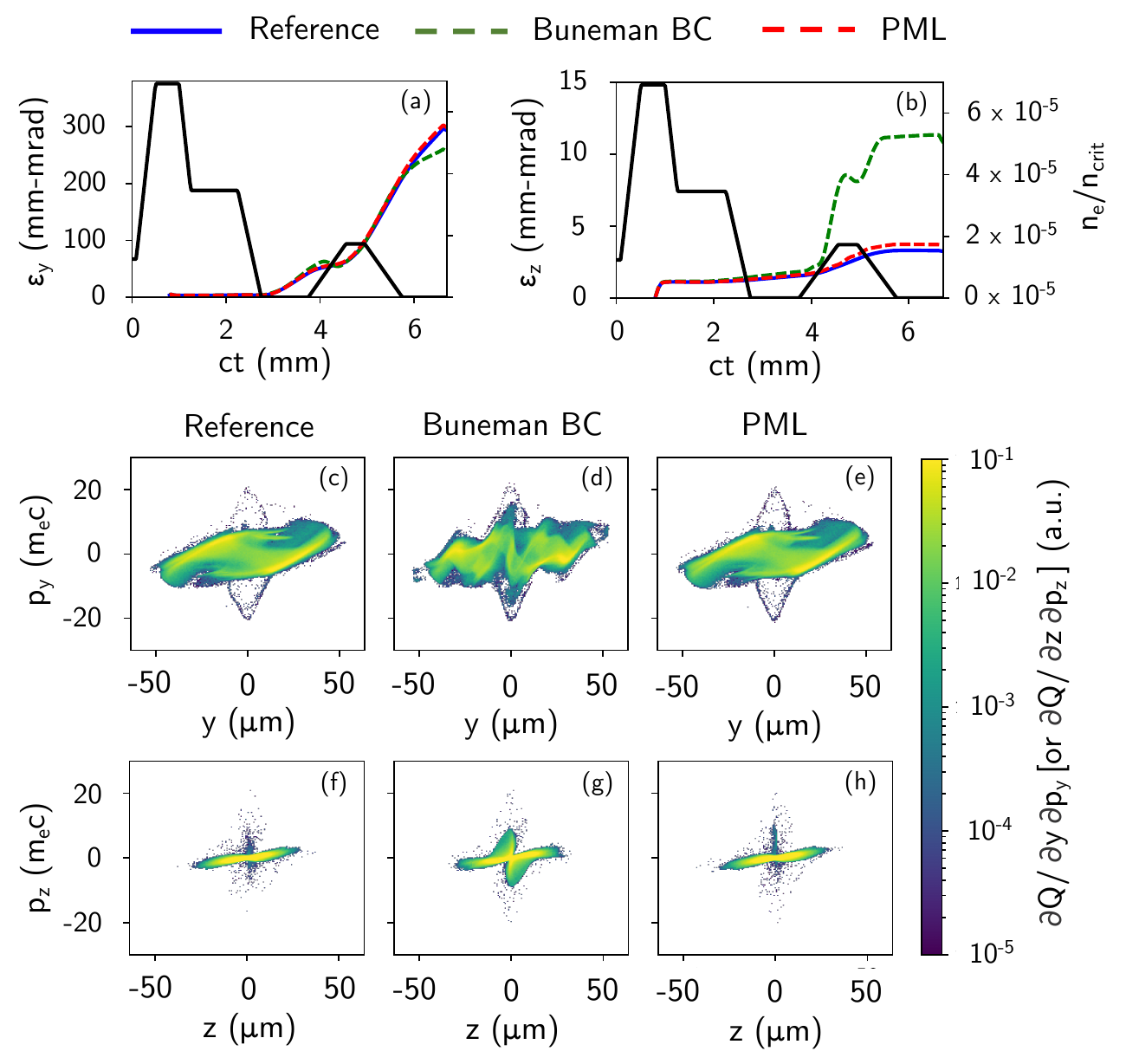}

  \caption{
  Results of Reference, Buneman BC and PML simulations. Panel a): evolution of the electron beam normalized emittance in the $y$ direction. Panel b): evolution of the electron beam normalized emittance in the $z$ direction. In both panels, The black line shows the plasma density profile.
  Panels c), d), e): electron beam distribution in the transverse phase space $y-p_y$. Panels f), g), h): electron beam distribution in the transverse phase space $z-p_z$. In these six panels, the beam is shown after 4.75 mm of propagation. All the panels count only the electrons obtained from ionization and with energy greater than 145 MeV.}\label{fig:2d_am_std_emittance}
\end{figure}

But the benefits of PML are even clearer on Fig. \ref{fig:2d_am_std_emittance} which shows the beam's normalized emittance evolution and transverse phase space distribution at 4.75 mm of propagation.
The transverse normalized emittance of the beam is defined as $\varepsilon_{i}= (m_ec)^{-1}\sqrt{\sigma_i^2\sigma_{p_i}^2-\sigma_{i-p_i}^2}$, with $i={y,z}$, where $\sigma_i$, $\sigma_{p_i}$, $\sigma_{i-p_i}$ are respectively the electron beam rms width in the $i$ coordinate, the rms width of the momentum component $p_i$ and the correlation between the coordinate $i$ and the momentum component $p_i$ \cite{Ferrario2016}.
This quantity is particularly important to evaluate an electron beam's suitability for applications, where typically the lowest values are more desirable.

As before, the Reference and PML simulation are in good agreement, while the Buneman BC simulation shows an explosion of the normalized emittance in the $z$ direction, due to the effects on the plasma of the laser reflections at the boundaries.

The Reference simulation ran with a 512 MPI processes and 20 OpenMP threads configuration for a total of 126 k core.hours.
Both the Buneman BC and PML simulations ran on a lower number of compute nodes  with only 128 MPI processes and 20 OpenMP threads each, for a total of 38.1 k core.hours and 38.5 k core.hours respectively.
The Buneman BC and PML simulations cost a comparable amount of computing resources, 3.3 less than the Reference simulation, which highlights the need for efficient absorbing boundary conditions.
Furthermore, the PML simulation used essentially the same amount of resources as the Buneman BC simulation, but yielded results that are considerably more similar to the Reference simulation by preventing laser reflections at the borders. 

\section{PML for the second order wave equation}\label{sec_wave_propagation}

This section describes how the PML medium can also be used as ABC for the full envelope wave equation and how it has been implemented in \Smilei.
The motivation for this work is that PIC codes are capable of simulating a laser propagation in plasma
via resolution of the wave equation obeyed by its envelope through an efficient, highly parallelizable explicit solver \cite{Terzani2019} for which no satisfactory ABC exist today to the knowledge of the authors.
This section therefore focuses on solutions to the envelope model and explains how to derive PML for this kind of equations.
The setup of the PML medium for a second order equation is more complicated than it is for the simpler first order Maxwell's equations and presents additional limitations that are discussed.

\subsection{Envelope Equation in Smilei}
In the envelope model formulation of laser-plasma interaction, the transverse component $\hat{A}$ of the vector potential of a laser pulse oscillating at frequency $\omega_0=k_0c$ and propagating in the $x$ direction is assumed to have the form: 
\begin{equation}
\hat{A}(\mathbf{r},t)=\textrm{Re}\left[\tilde{A}(\mathbf{r},t)e^{ik_0(x-ct)}\right],
\label{eq:envelope_definition}
\end{equation}
where $\tilde{A}$, the low frequency part of $\hat{A}$, is referred to as the complex envelope of the transverse component of the laser's vector potential.
With this ansatz, if the characteristic space-time scales of the laser-pulse and of the plasma wave it excites are comparable or larger than $k_0^{-1}$ and $\omega_0^{-1}$, the equations describing laser-plasma interaction can be reformulated in terms of low-frequency fields only \cite{Mora1997,Cowan2011}. This envelope, also called ponderomotive guiding center, formulation of laser-plasma interaction allows considerable savings in PIC simulations because it relaxes constraints on the required resolutions in both time and space \cite{Mora1997,Cowan2011,Gordon2000,Huang2006,Benedetti2010,Tomassini2016MatchingSF,Terzani2019,Massimo2019_3D,Massimo2019cylindrical,Silva2019,Massimo2020ionization,Terzani2021}.

The propagation of the vector potential $\hat{A}$ in vacuum satisfies the wave equation
%
%
\begin{equation}
\nabla^2 \hat{A} - \dfrac{1}{c^2}\partielsnddouble{\hat{A}}{t} =0.
\label{eq:wave_eq}
\end{equation}
By introducing the envelope definition in Eq.\ref{eq:envelope_definition} into the wave equation in Eq.\ref{eq:wave_eq}, the following envelope equation is obtained \cite{Terzani2019}
\begin{equation}
\nabla^2 \tilde{A} +2ik_0\left(\partiel{\tilde{A}}{x} +\dfrac{1}{c}\partiel{\tilde{A}}{t}\right) - \dfrac{1}{c^2}\partielsnddouble{\tilde{A}}{t} =0.
\label{eq:envelope_equation}
\end{equation}

This is the equation eventually discretized and solved in the code and for which a PML is derived in the following section. Note that a given $(\omega, k_x,k_y,k_z)$ component of $\hat{A}$ in this model is described by a low frequency $(\tilde{\omega}=\omega-\omega_0, \tilde{k_x}=k_x-k_0,k_y,k_z)$ component of $\tilde{A}$.

It is important to note that the longitudinal direction $x$ is different from the others since its wave vector component is shifted whereas, in the other directions, they are not.
As a result, deriving a PML in this particular direction raises particular stability issues and is left out of the scope of this paper.
In the next section, the derivation of transverse PML, sufficient for most of the simulations with a propagating laser pulse, are presented.



\subsection{Transverse PML for the envelope wave equation in Cartesian geometry}

This section presents how to derive the envelope propagation equation in transverse PML ($y$ or $z$-PML in the Cartesian geometry as represented on Fig.\ref{fig:pml-scheme}) and how it can be solved with an FDTD scheme.
The envelope propagation equation in vacuum in Eq.\ref{eq:envelope_equation} can be adjusted to the PML media using stretched coordinates as in section \ref{sec:stretched_coordinates} which gives
\begin{equation} 
\partielsnddouble{\tilde{A}}{\tilde{x}}+2ik_0\left(\partiel{\tilde{A}}{\tilde{x}} + \dfrac{1}{c}\partiel{\tilde{A}}{t}\right)+\partielsnddouble{\tilde{A}}{\tilde{y}}+\partielsnddouble{\tilde{A}}{\tilde{z}}-\dfrac{1}{c^2}\partielsnddouble{\tilde{A}}{t} = 0,
\end{equation}
and using the definitions in Eq. \ref{eq:variable_transform}
\begin{equation} 
\dfrac{1}{s_x}\partiel{}{x}\left(\dfrac{1}{s_x}\partiel{\tilde{A}}{x}\right) + 2ik_0\left(\frac{1}{s_x}\partiel{\tilde{A}}{x} +\dfrac{1}{c}\partiel{\tilde{A}}{t}\right) + \dfrac{1}{s_y}\partiel{}{y}\left(\dfrac{1}{s_y}\partiel{\tilde{A}}{y}\right) + \dfrac{1}{s_z}\partiel{}{z}\left(\dfrac{1}{s_z}\partiel{\tilde{A}}{z}\right) -\dfrac{1}{c^2}\partielsnddouble{\tilde{A}}{t} = 0.
\label{eq:snd-order-pml-eq-0}
\end{equation}
%
Let's consider for instance a laser envelope $\tilde{A}$ propagating in a transverse y-PML.
In that case, and outside of a corner, we have $s_x=s_z=1$ and equation (\ref{eq:snd-order-pml-eq-0}) becomes
%
\begin{equation}
-\dfrac{s_y'}{s_y^3}\partiel{\tilde{A}}{y} + \left(\dfrac{1}{s_y^2}-1\right) \partielsnddouble{\tilde{A}}{y}  + 2ik_0\left(\partiel{\tilde{A}}{x} +\dfrac{1}{c}\partiel{\tilde{A}}{t}\right)+ \partielsnddouble{\tilde{A}}{x}+\partielsnddouble{\tilde{A}}{y}+\partielsnddouble{\tilde{A}}{z}-\dfrac{1}{c^2}\partielsnddouble{\tilde{A}}{t} = 0,
\label{eq:wave-pml-y}
\end{equation}
where $s_y'=\dfrac{ds_y}{dy}$.

The standard wave equation is recovered with several additional terms corresponding to the effect of the PML on the envelope.
The frequency components of $\tilde{A}$, are centered around $\omega=0$ which makes the complex susceptibility in Eq.\ref{eq:susceptibility} not well defined.
In addition, it is also observed that evanescent waves and waves with grazing incidence are poorly absorbed by PML and accumulate numerical dispersion which can lead to instabilities over long simulation times.
In order to stabilize the PML and improve absorption for such waves, the susceptibility of the PML medium for the envelope is modified by a complex frequency shift of the form \cite{Roden2000,kuzuoglu1996,Martin2007}
\begin{equation}
s_y(y) = \varepsilon_y(y) + \dfrac{\sigma_y(y)}{\alpha_y(y)+i\omega\varepsilon_0},
\end{equation}
where a real valued $\alpha$ parameter, function of the normal position $y$ inside the PML, has been added to the denominator. $s_y$ can be decomposed into its real and imaginary parts
\begin{equation}
s_y=\varepsilon_y+\frac{\alpha\sigma}{\alpha^2+(\omega\varepsilon_0)^2}-\frac{i\omega\varepsilon_0\sigma}{\alpha^2+(\omega\varepsilon_0)^2}.
\end{equation}

The second term contributes to damping low frequency evanescent waves when it is positive so as long as $\alpha$ is of the same sign as $\sigma$.
Note that the damping of non evanescent propagative waves, driven by the imaginary part of $s_y$, now also becomes frequency dependent.
Lower frequencies are not efficiently damped anymore and only frequencies such as $\omega\epsilon_0>\alpha$ are properly damped.
Therefore the value of the $\alpha$ parameter must be chosen as the result of a trade-of between those two features: small enough to damp most propagating waves and large enough to damp low frequency evanescent waves.
A common solution is to choose a linear profile of $\alpha$: maximal at the PML interface with the physical domain and zero at the PML extremity \cite{Roden2000,Martin2007}.
But it also possible to keep $\alpha$ constant at a sufficiently low value \cite{Roden2000}.
This second solution seems appropriate for a simulation with an envelope model (see Section \ref{sec_pml_envelope_benchmarks}) and is the one used in \Smilei.

%

%
We can now evaluate
\begin{align}
\dfrac{s_y'}{s_y^3} =\varepsilon_y'\dfrac{(i\omega\varepsilon_0+\alpha_y)^3}{(\varepsilon_y i\omega\varepsilon_0 +\varepsilon_y\alpha_y+\sigma_y)^3} + \dfrac{\big(\sigma_y'(\alpha_y+i\omega\varepsilon_0)-\sigma_y\alpha_y'\big)\big( i\omega\varepsilon_0+\alpha_y \big)}{(\varepsilon_y i\omega\varepsilon_0 + \varepsilon_y\alpha_y+\sigma_y)^3}.
\end{align}
We assign $p = \varepsilon_y i\omega\varepsilon_0 + \varepsilon_y\alpha_y+\sigma_y$ to obtain
\begin{align}
\dfrac{1}{s_y} &= \dfrac{p-\sigma_y}{\varepsilon_y p}\label{eq:one-over-s},\\[12pt]
%
%
\dfrac{s_y'}{s_y^3} &=  \varepsilon_y' \dfrac{p^3-3p^2\sigma_y+3p\sigma_y^2 -\sigma_y^3}{\varepsilon_y^3 p^3} + \dfrac{\sigma_y'p^2-2p\sigma_y\sigma_y'+\sigma_y^2\sigma_y' - \varepsilon_y\sigma_y\alpha_y' p+\varepsilon_y\sigma_y^2\alpha_y'}{\varepsilon_y^2 p^3}.
\end{align}
We can rewrite this last expression with a partial fraction decomposition
\begin{align}
\dfrac{s_y'}{s_y^3} = \dfrac{1}{\varepsilon_y^3}\big(\varepsilon_y'+\dfrac{\varepsilon_y\sigma_y'-3\sigma_y\varepsilon_y'}{p}+\dfrac{3\sigma_y^2\varepsilon_y'-2\varepsilon_y\sigma_y\sigma_y'-\varepsilon_y^2\sigma_y\alpha_y'}{p^2}+\dfrac{\varepsilon_y\sigma_y^2\sigma_y'+\varepsilon_y^2\sigma_y^2\alpha_y'-\varepsilon_y'\sigma_y^3}{p^3} \big) \label{eq:s-prime-over-s-cube}.
\end{align}
Combining Eq.\ref{eq:wave-pml-y}, Eq.\ref{eq:one-over-s} and Eq.\ref{eq:s-prime-over-s-cube}, we obtain the wave-PML propagation equation where terms in the power of $p^{-1}$ are well identified
\begin{align}
\frac{1}{c^2}\partielsnddouble{\tilde{A}}{t} - 2 i k_0 \dfrac{1}{c}\partiel{\tilde{A}}{t} =
&\partielsnddouble{\tilde{A}}{x} + 2 i k_0 \partiel{\tilde{A}}{x} + \partielsnddouble{\tilde{A}}{y}+\partielsnddouble{\tilde{A}}{z} \nonumber \\[12pt]
&+ \big(\dfrac{1}{\varepsilon_y^2} - 1 \big)\partielsnddouble{\tilde{A}}{y} -\dfrac{\varepsilon_y'}{\varepsilon_y^3}\partiel{\tilde{A}}{y}\nonumber\\[12pt]
&+ \frac{1}{p}\left(\dfrac{3\sigma_y\varepsilon_y'-\varepsilon_y\sigma_y'}{\varepsilon_y^3 }\partiel{\tilde{A}}{y}-\dfrac{2\sigma_y}{\varepsilon_y^2 }\partielsnddouble{\tilde{A}}{y}\right)\nonumber\\[12pt]
&+\frac{1}{p^2} \left(\dfrac{2\varepsilon_y\sigma_y\sigma_y'+\varepsilon_y^2\sigma_y\alpha_y'-3\sigma_y^2\varepsilon_y'}{\varepsilon_y^3}\partiel{\tilde{A}}{y}+\dfrac{\sigma_y^2}{\varepsilon_y^2 }\partielsnddouble{\tilde{A}}{y}\right)\nonumber\\[12pt]
&+\frac{1}{p^3}\dfrac{\varepsilon_y'\sigma_y^3-\varepsilon_y\sigma_y^2\sigma_y'-\varepsilon_y^2\sigma_y^2\alpha_y'}{\varepsilon_y^3}\partiel{\tilde{A}}{y}.
\end{align}
Note that only $p$ contains $i\omega$ terms. 
In order to avoid division by $i\omega$ which would translate into inconvenient time integrations when implementing the FDTD scheme, auxiliary variables $u_{0,x},u_{1,x},u_{2,x},u_{3,x}$ are introduced in order to avoid having $p$ as a denominator
\begin{equation}
u_{0,y} = \big(\dfrac{1}{\varepsilon_y^2} - 1 \big)\partielsnddouble{\tilde{A}}{y} -\dfrac{\varepsilon_y'}{\varepsilon_y^3}\partiel{\tilde{A}}{y} + \tilde u_{1,y},
\end{equation}

\begin{align}
p \tilde u_{1,y} &=\dfrac{3\sigma_y\varepsilon_y'-\varepsilon_y\sigma_y'}{\varepsilon_y^3}\partiel{\tilde{A}}{y}-\dfrac{2\sigma_y}{\varepsilon_y^2}\partielsnddouble{\tilde{A}}{y} + \tilde u_{2,y}, \\[12pt]
p \tilde u_{2,y} &= \dfrac{2\varepsilon_y\sigma_y\sigma_y'+\varepsilon_y^2\sigma_y\alpha_y'-3\sigma_y^2\varepsilon_y'}{\varepsilon_y^3}\partiel{\tilde{A}}{y}+\dfrac{\sigma_y^2}{\varepsilon_y^2}\partielsnddouble{\tilde{A}}{y} + \tilde u_{3,y}, \\[12pt]
p \tilde u_{3,y} &= \dfrac{\varepsilon_y'\sigma_y^3-\varepsilon_y\sigma_y^2\sigma_y'-\varepsilon_y^2\sigma_y^2\alpha_y'}{\varepsilon_y^3}\partiel{\tilde{A}}{y}.
\end{align}
In this system, all $\tilde u_{i,y}$, $i\in[0,3]$, are functions of $\tilde{A}$ and $\tilde u_{i+1,y}$ except $\tilde u_{3,y}$ which depends only on $\tilde{A}$.
These, so called, Auxiliary Differential Equations (ADE) can therefore be resolved explicitly in a sequential manner.
In the present form, ADE can be easily discretized using any FDTD scheme.
Using the envelope assumption again, the auxiliary variables $\tilde u_{i,y}$ have the same form as $\tilde{A}$ and their high frequency counterparts $\hat{u}_{i,y}$ can be derived as follows
\begin{align}
\frac{\partial \hat{u}_{i,y}}{\partial t} = -i\omega \hat{u}_{i,y} &= \frac{\partial \tilde{u}_{i,y}}{\partial t} \e{i(k_0 x - \omega_0 t)} - i\omega_0 \tilde{u}_{i,y} \e{i(k_0 x - \omega_0 t)}\nonumber\\
&= -i(\omega-\omega_0) \tilde{u}_{i,y} \e{i(k_0 x - \omega_0 t)} - i\omega_0 \tilde{u}_{i,y} \e{i(k_0 x - \omega_0 t)}.
\end{align}
From the last equality it is observed that
\begin{equation}\label{eq:envelope_time_derivative}
i\omega  \tilde{u}_{i,y} = i\omega_0 \tilde{u}_{i,y} - \frac{\partial \tilde{u}_{i,y}}{\partial t},
\end{equation}
which can be introduced in the expression of $p$
\begin{align}
\partielsnddouble{\tilde{A}}{t} - 2 i \omega_0 \partiel{\tilde{A}}{t} &= \partielsnddouble{\tilde{A}}{x} + 2 i k_0 \partiel{\tilde{A}}{x} + \partielsnddouble{\tilde{A}}{y}+\partielsnddouble{\tilde{A}}{z} + u_{0,y} \label{eq:ade-u0}\\[12pt]
u_{0,y} &= \big(\dfrac{1}{\varepsilon_y^2} - 1 \big)\partielsnddouble{\tilde{A}}{y} -\dfrac{\varepsilon_y'}{\varepsilon_y^3}\partiel{\tilde{A}}{y} + \tilde u_{1,y} \\[12pt] 
(\partial_t - i \omega_0 - \alpha_y/\varepsilon_0 - \sigma_y/\varepsilon_0\varepsilon_y) \tilde u_{1,y}&=-\dfrac{3\sigma_y\varepsilon_y'-\varepsilon_y\sigma_y'}{\varepsilon_0\varepsilon_y^4}\partiel{\tilde{A}}{y}+\dfrac{2\sigma_y}{\varepsilon_0\varepsilon_y^3}\partielsnddouble{\tilde{A}}{y} - \dfrac{\tilde u_{2,y}}{\varepsilon_0\varepsilon_y} \label{eq:ade-u1}\\[12pt]
(\partial_t - i \omega_0 - \alpha_y/\varepsilon_0 - \sigma_y/\varepsilon_0\varepsilon_y) \tilde u_{2,y}&= -\dfrac{2\varepsilon_y\sigma_y\sigma_y'+\varepsilon_y^2\sigma_y\alpha_y'-3\sigma_y^2\varepsilon_y'}{\varepsilon_0\varepsilon_y^4}\partiel{\tilde{A}}{y}-\dfrac{\sigma_y^2}{\varepsilon_0\varepsilon_y^3}\partielsnddouble{\tilde{A}}{y} - \dfrac{\tilde u_{3,y}}{\varepsilon_0\varepsilon_y} \label{eq:ade-u2}\\[12pt]
(\partial_t - i \omega_0 - \alpha_y/\varepsilon_0 - \sigma_y/\varepsilon_0\varepsilon_y) \tilde u_{3,y}&= -\dfrac{\varepsilon_y'\sigma_y^3-\varepsilon_y\sigma_y^2\sigma_y'-\varepsilon_y^2\sigma_y^2\alpha_y'}{\varepsilon_0\varepsilon_y^4}\partiel{\tilde{A}}{y}\label{eq:ade-u3}.
\end{align}
%
%
This set of auxiliary  equations can be solved sequentially.
Starting with $\tilde u_{3,y}$ and using the central finite difference staggered in time 
, Eq.\ref{eq:ade-u3} is discretized as follows
%
\begin{align}\label{eq:u3y}
\tilde{u}_{3,y}^{n+1/2}
=& \dfrac{ 2 + \Delta t (i\omega_0 + \alpha_y/\varepsilon_0 + \sigma_y/\varepsilon_0\varepsilon_y)}{ 2 - \Delta t (i\omega_0 + \alpha_y/\varepsilon_0 + \sigma_y/\varepsilon_0\varepsilon_y)}\tilde{u}_{3,y}^{n-1/2} \nonumber \\
&- \dfrac{2\Delta t}{ 2 - \Delta t (i\omega_0 + \alpha_y/\varepsilon_0 + \sigma_y/\varepsilon_0\varepsilon_y )} \dfrac{\varepsilon_y'\sigma_y^3-\varepsilon_y\sigma_y^2\sigma_y'-\varepsilon_y^2\sigma_y^2\alpha_y'}{\varepsilon_0\varepsilon_y^4}\partiel{\tilde{A}}{y}
\end{align}
With $\tilde u_{3,y}$ known by using Eq.\ref{eq:u3y}, it is possible to determine $\tilde u_{2,y}$ via Eq. \ref{eq:ade-u2} and so on sequentially until eventually the whole ADE system is solved and the value of $\tilde{A}$ at time $n+1$ is evaluated via Eq. \ref{eq:ade-u0}.

\subsection{PML for the envelope wave equation in cylindrical geometry}

This section presents propagation equations for the envelope in a PML medium in the AM geometry.
As in the Cartesian case, only the transverse direction is treated. 
Therefore, in this geometry the discussion is limited to the implementation of the radial PML for the envelope model.
The envelope is always assumed to have cylindrical symmetry ($\partial_\theta\tilde{A}=0$) and thus to be entirely described by the mode $m=0$.
All derivatives of $\hat{A}$ along the $\theta$ variable are ignored.

The wave equation in vacuum in cylindrical symmetry reads \cite{Massimo2019cylindrical}:
\begin{equation}\label{envelope_equation}
\partial^2_x \hat{A}+ \frac{1}{r}\partial_r\left(r\partial_r \hat{A} \right)-\frac{1}{c^2}\partial^2_t\hat{A}=0,
\end{equation}
%
 which can be reformulated as
\begin{equation}
\partial^2_x \hat{G} + \partial^2_r \hat{G} - \partial_r \hat{A} -\frac{1}{c^2}\partial^2_t \hat{G}=0,
\end{equation}
where $\hat{G}=r\hat{A}$.
This formulation allows to avoid the $1/r$ dependence, and easily use stretched coordinates with any kind of FDTD scheme.
This comes at the cost of an additional step, with respect to the Cartesian implementation, since $\hat{G}$ needs to be computed prior to the evaluation of $\hat{A}$.
As in the Cartesian case, an ADE-scheme is used to evaluate the time derivative of $\tilde{G}$
\begin{align}
\partielsnddouble{\tilde{G}}{t} - 2 i \omega_0 \partiel{\tilde{G}}{t} &= \partielsnddouble{\tilde{G}}{x} + 2 i k_0 \partiel{\tilde{G}}{x} + \partielsnddouble{\tilde{G}}{r} - \partiel{\tilde{A}}{r} + u_{0,r} \label{eq:ade-u0-am}\\[12pt]
u_{0,r} &= \big(\dfrac{1}{\varepsilon_r^2} - 1 \big)\partielsnddouble{\tilde{G}}{r} -\dfrac{\varepsilon_r'}{\varepsilon_r^3}\partiel{\tilde{G}}{r} + (1-\dfrac{1}{\varepsilon_r})\partiel{\tilde{A}}{r} + \tilde u_{1,r} \\[12pt]
(\partial_t - i \omega_0 - \alpha_r/\varepsilon_0 - \sigma_r/\varepsilon_0\varepsilon_r) \tilde u_{1,r}&=-\dfrac{3\sigma_r\varepsilon_r'-\varepsilon_r\sigma_r'}{\varepsilon_0\varepsilon_r^4}\partiel{\tilde{G}}{r}+\dfrac{2\sigma_r}{\varepsilon_0\varepsilon_r^3}\partielsnddouble{\tilde{G}}{r} - \dfrac{\sigma_r}{\varepsilon_r^2}\partiel{\tilde{A}}{r} - \dfrac{\tilde u_{2,r}}{\varepsilon_0\varepsilon_r} \label{eq:ade-u1-am}\\[12pt]
(\partial_t - i \omega_0 - \alpha_r/\varepsilon_0 - \sigma_r/\varepsilon_0\varepsilon_r) \tilde u_{2,r}&= -\dfrac{2\varepsilon_r\sigma_r\sigma_r'+\varepsilon_r^2\sigma_r\alpha_r'-3\sigma_r^2\varepsilon_r'}{\varepsilon_0\varepsilon_r^4}\partiel{\tilde{G}}{r}-\dfrac{\sigma_r^2}{\varepsilon_0\varepsilon_r^3}\partielsnddouble{\tilde{G}}{r} - \dfrac{\tilde u_{3,r}}{\varepsilon_0\varepsilon_r} \label{eq:ade-u2-am}\\[12pt]
(\partial_t - i \omega_0 - \alpha_r/\varepsilon_0 - \sigma_r/\varepsilon_0\varepsilon_r) \tilde u_{3,r}&= -\dfrac{\varepsilon_r'\sigma_r^3-\varepsilon_r\sigma_r^2\sigma_r'-\varepsilon_r^2\sigma_r^2\alpha_r'}{\varepsilon_0\varepsilon_r^4}\partiel{\tilde{G}}{r}.\label{eq:ade-u3-am}
\end{align}
The resolution of this system can be treated as in the Cartesian case.
Finally, in order to recover $\tilde{A}$ from $\tilde{G}$, one has to recall the stretched coordinate $\tilde{r}$ along $r$
\begin{equation}
    \tilde{r}(r) = \int_{r_0}^{r} \varepsilon_r(r') + \dfrac{\sigma_r(r')}{\alpha_r(r') + i\omega} \diff r' = r0 + E (r) + \dfrac{\Sigma(r)}{\alpha_r + i\omega},
\end{equation}
where $\alpha_r$ is chosen constant along $r$ as explained in the previous section.
$\Sigma$ and $E$ are respectively the integrals of $\sigma$ and $\varepsilon$ over the PML radial direction.
Writing $\tilde{G}=\tilde{r}\tilde{A}$ gives
\begin{equation}
\left(i\omega + \alpha_r + \dfrac{\Sigma(r)}{ r0 + E (r) }\right)\tilde{A} = \dfrac{i\omega + \alpha_r}{ r0 + E (r)}\tilde{G}.
\end{equation}
Recalling Eq. \ref{eq:envelope_time_derivative}, the time derivative of $\tilde{A}$ can be introduced
\begin{equation}
\partial_t \tilde{A} - \left(i\omega_0 + \alpha_r + \dfrac{\Sigma(r)}{ r0 + E (r) }\right)\tilde{A} = \dfrac{\partial_t \tilde{G} - (i\omega_0 + \alpha_r)\tilde{G}}{ r0 + E (r)}.
\end{equation}
Since  $\tilde{A}$ and  $\tilde{G}$ are evaluated at the same times and positions on the grid, an FDTD scheme finally yields
%
\begin{align}
\tilde{A}^{n+1} = &
\tilde{A}^{n-1}
\dfrac{1+\Delta t\left(i\omega_0 + \alpha_r + \dfrac{\Sigma(r)}{ r0 + E (r) }\right)}{1-\Delta t\left(i\omega_0 + \alpha_r + \dfrac{\Sigma(r)}{ r0 + E (r) }\right)} \nonumber \\
&+ \dfrac{1}{(r0 + E (r))}\dfrac{\left[ \tilde{G}^{n+1}(1-\Delta t(i\omega_0 + \alpha_r))-\tilde{G}^{n-1}(1+\Delta t(i\omega_0 + \alpha_r)\right]}{ 1-\Delta t\left(i\omega_0 + \alpha_r + \dfrac{\Sigma(r)}{ r0 + E (r) }\right)}.
\end{align}

This final expression gives the evolution in time of the envelope field in the PML as implemented in \Smilei.

\subsection{PML envelope benchmarks in AM geometry}\label{sec_pml_envelope_benchmarks}

As benchmarks for the PML boundary conditions for the envelope equation, in this section the results of three simulations in cylindrical geometry with a single azimuthal mode are shown.
The physical case is identical to the first benchmark presented in section \ref{sec:benchmark_standard_cylindrical_PML}. 
The mesh and timestep are also the same, except for the domain's transverse size for the so called "reflective" and "PML" simulations, set to $L_r=416$ $c/\omega_0$. A preliminary clarification is that although the physical setup is the same, the results of the simulations using an envelope model presented in this section present differences with respect to the results of the simulations not using an envelope model shown in section \ref{sec:benchmark_standard_cylindrical_PML}.
It has been shown that at convergence the two models (with and without envelope) yield the same results \cite{Terzani2021}, but a particularly small $\Delta x$ and $\Delta t$ is necessary in the standard simulations to reach convergence.
Furthermore, the ionization models used with the envelope model are not the same (their differences are discussed in \cite{Massimo2020ionization}) and can yield different results when the simulations are not at convergence.
A full study of the convergence of the two models is well beyond the scope of this article which only aims at illustrating how the PML can significantly cope with the reflections of the complex envelope at the simulation boundary. 


The three simulations presented in this section are the "Reference" simulation, which has a transversely large physical domain with reflective boundary conditions for the envelope equation, the "Reflective BC" simulation which has a transversely smaller physical domain with the same reflective boundary conditions and the "PML" simulation which also has a small physical domain but uses PML boundary conditions.
It is important to note that, to the authors knowledge, there are no other boundary conditions described in the literature for this particular form of the envelope equation, which is the reason why reflective boundary conditions are used.
For the electromagnetic fields, the transverse boundary conditions are ABC of type Buneman for the "reference" and "reflective" simulations, and PML for the boundaries of the "PML" simulation.

\begin{figure}[h!]
    \includegraphics[width=\textwidth]{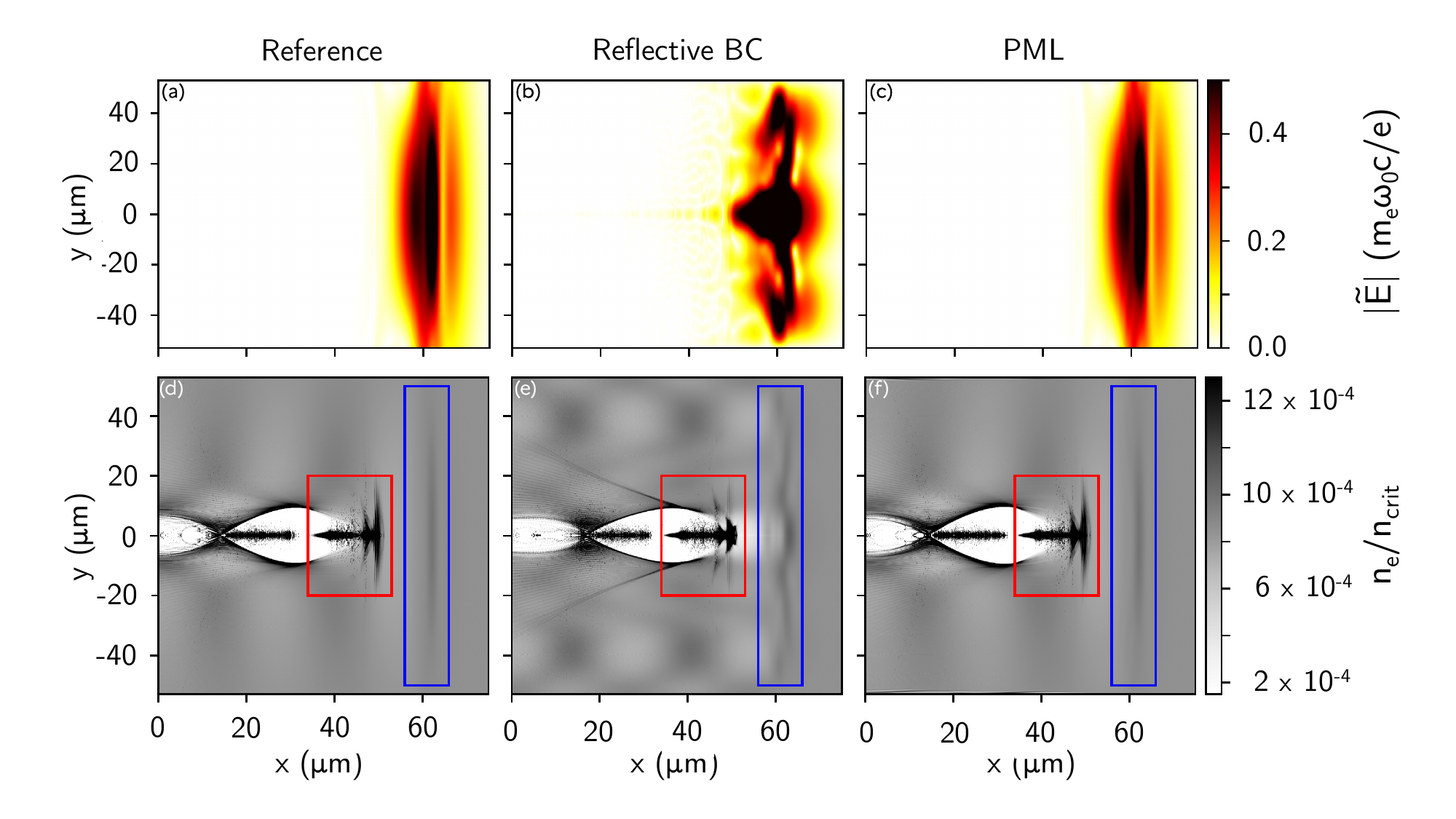}

  \caption{Results of Reference, Reflective BC and PML simulations with the laser envelope model in the AM geometry after 4.75 mm of propagation. Top panels: $|\tilde{E}|$ field on the plane xy; Bottom panels: electron density on the plane xy.
  Panels a),d): Reference" simulation; Panels b),e): Reflective BC simulation; Panels c),f) PML simulation. For readers' convenience, a blue rectangle was added in the zone where the peak field of the laser pulse lies; similarly, a red rectangle was added to mark the position of the main accelerated electron beam. }\label{fig:2d_am_env}
\end{figure}

Figure \ref{fig:2d_am_env} shows the envelope of the laser transverse electric field $|\tilde{E}|$ on the $xy$ plane after 4.75 mm of propagation in the three simulations, in a window close to the laser pulse. Results similar to those in Fig. \ref{fig:Bz_and_density_cylindrical_standard} can be observed.
As in Fig. \ref{fig:Bz_and_density_cylindrical_standard}, the simulation with PML boundary conditions shows results very similar to those of the "reference" simulation, while the "reflective" simulation shows artificial reflections at the borders, creating an interference pattern for the laser field, with consequences on the wake density distribution.

\begin{figure}[h!]
    \includegraphics[width=0.9\textwidth]{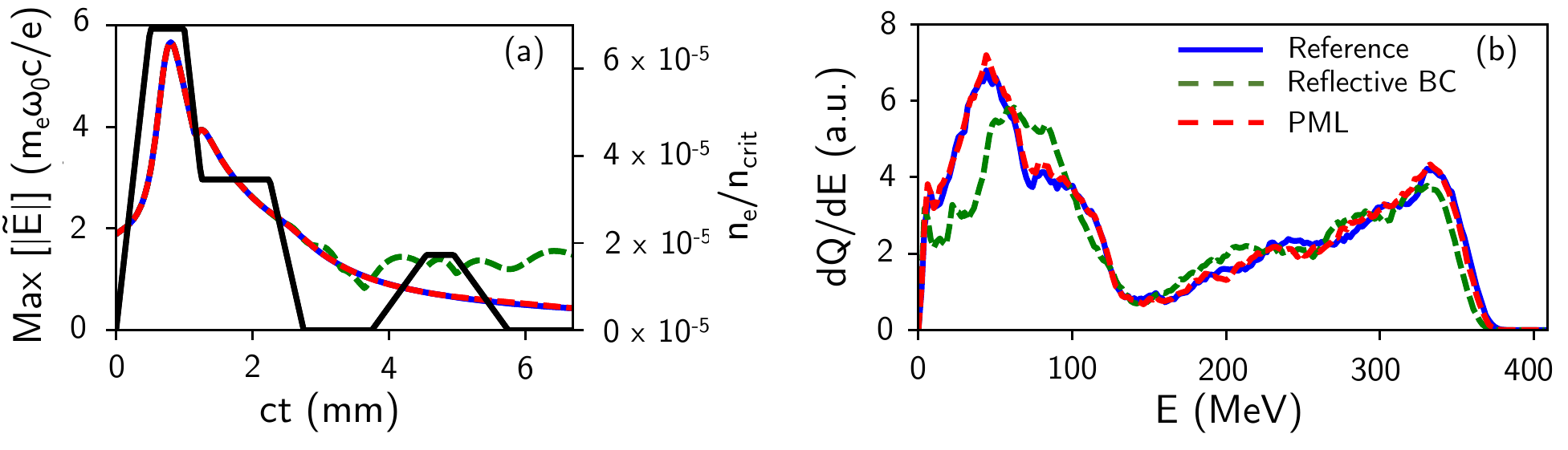}

  \caption{Results of "reference", "reflective" and "PML" simulations with the laser envelope model. Panel a): evolution of the peak laser field $max(|\tilde{E}|)$. The black line shows the plasma density profile.
  Panel b): electron spectra obtained in the three simulations after 4.75 mm of propagation. Only the electrons obtained from ionization are shown }\label{fig:2d_am_env_evo_temp}
\end{figure}

Figure \ref{fig:2d_am_env_evo_temp}, as Fig. \ref{fig:2d_am_std_emittance}, shows the evolution of the peak absolute value of the laser's envelope of the transverse electric field and the spectra of the electrons obtained from ionization in the three simulations.
The reflections in the laser field after a long distance (Fig.\ref{fig:2d_am_env_evo_temp}a) are particularly striking, due to the use of reflective boundary conditions, while the "reference" and "PML" simulations are in agreement.
These reflections in the "reflective" simulation seem to have an effect on both the low energy and high energy electrons (Panel b).

\begin{figure}[h!]
    \includegraphics[width=0.9\textwidth]{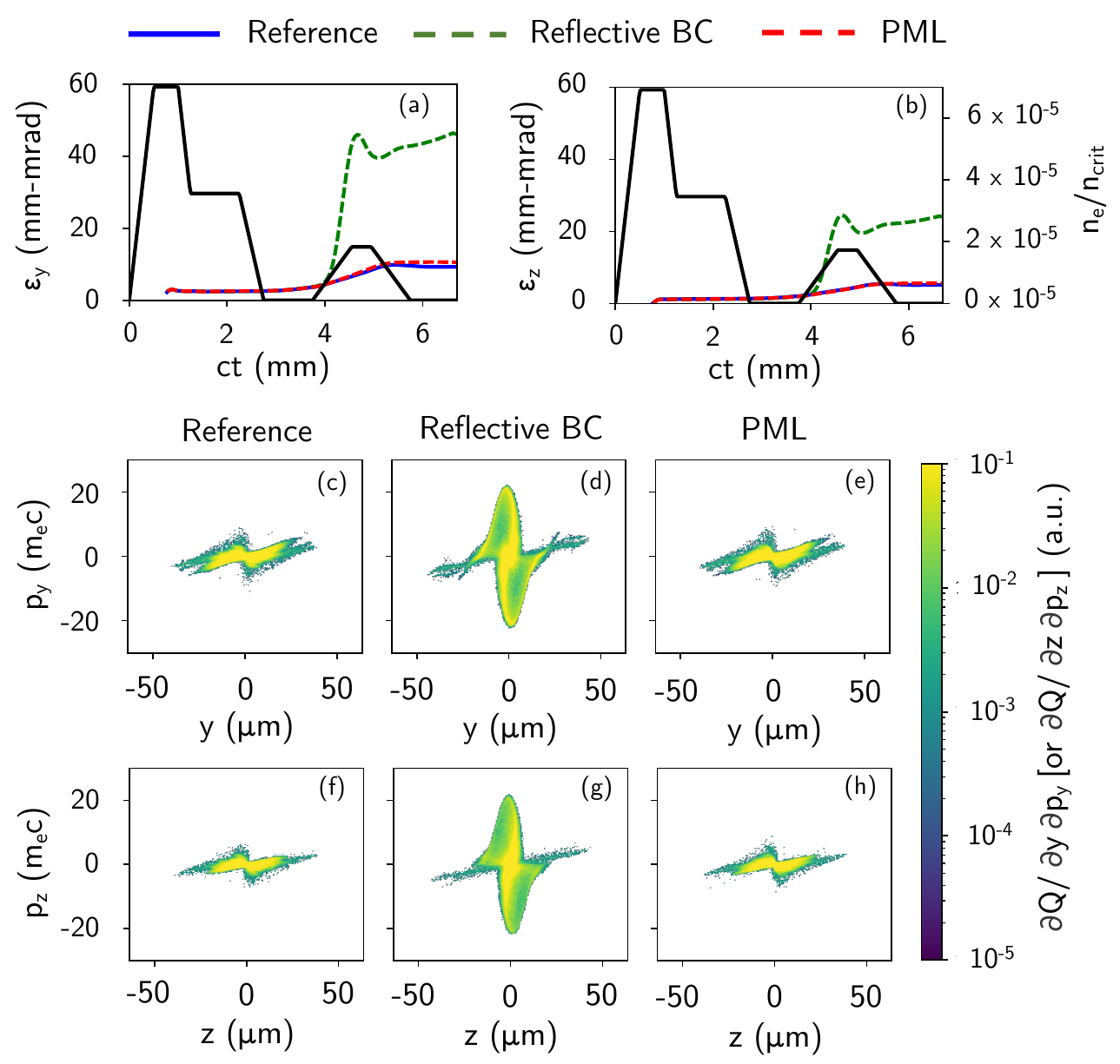}

  \caption{Results of Reference, Reflective BC and PML simulations with the laser envelope model. Panel a): evolution of the electron beam normalized emittance in the $y$ direction. Panel b): evolution of the electron beam normalized emittance in the $z$ direction. In both panels, The black line shows the plasma density profile.
  Panels c), d), e): electron beam distribution in the transverse phase space $y-p_y$. Panels f), g), h): electron beam distribution in the transverse phase space $z-p_z$. In these six panels, the beam is shown after 4.75 mm of propagation. All the panels count only electrons obtained from ionization and with energy greater than 145 MeV.}\label{fig:2d_am_env_emittance}
\end{figure}

Figure \ref{fig:2d_am_env_emittance} shows the evolution of the beam's normalized emittance evolution and transverse phase space distribution after 4.75 mm of propagation. As before, the Reference and PML simulation are in good agreement, while the Reflective BC simulation shows an explosion of the emittance (even a factor 4) after a long distance, due to the effects on the plasma of the laser reflections at the borders.

The computing resources needed by these simulations should also be considered.
Using an envelope model significantly reduces the required computational resources with respect to a standard simulation \cite{Terzani2019,Massimo2019_3D,Massimo2020}.
It is the first motivation for using this reduced model hence the interest for designing proper open boundary conditions for the envelope propagation equation.
The Reference simulation ran with 28 MPI processes and 20 OpenMP threads, for a total of 8.4 kcpu-hours. The Reflective and PML simulations ran in the same configuration but for a total of 2.4 kcpu-hours and 2.6 kcpu-hours respectively, i.e. less than a third of the energy budget of the Reference simulation.
The PML simulation used an amount of resources similar to that used by the Reflective BC simulation, but gave results with accuracy comparable to the Reference simulation, which was three times as expensive. 

\section{Conclusions}

Since the capability of ABC to absorb outgoing waves remains limited, PML are proposed
in the literature as an improvement for the simulation of open boundaries.
They come with an additional computational cost but can easily be adjusted to the required accuracy level by tuning the number of PML cells and their absorbing properties and can be adapted to the desired geometry and numerical method.


In this paper, after reviewing the complex coordinates stretching derivation of the PML for Maxwell's equations in the time domain in both Cartesian and cylindrical geometries, PML have been derived for a cylindrical geometry with azimuthal Fourier decomposition, particularly useful for PIC simulations. Additionally and for the first time, a formulation of the PML in the time domain for the envelope equation has been presented. 
This result opens a broad range of new possible applications for the envelope model which saves a significant amount of computing time and energy whenever it can be used.

The general method used for the derivation of the PML has been detailed and the resulting numerical schemes have been implemented in the PIC code \Smilei.
Demonstrations of their excellent results both in accuracy and performance have been provided.

Nevertheless, PML, like most analytical ABC, are still theoretically limited to interfaces with vacuum.
Wave reflections can still be observed if PML are directly adjacent to plasma since they are assumed to be in vacuum and reflections on a plasma-vacuum interface are expected.
Future work will therefore include investigations to improve the PML-plasma interface.
Some leads on how to handle energetic outgoing particles already exist \cite{Copplestone2017,PML_lehe} and might be generalized.
Another limit of the PML lies in their lack of stability over very long simulation times when such interfaces are simulated.
Even for Maxwell's equations, instabilities are sometimes triggered eventually because of the error accumulation at the plasma-PML interfaces.
The complex frequency shift PML, already implemented for the envelope model, are a possible method to circumvent the rare occurrences of these long term instabilities and will be made available in a future work for standard fields as well.
The PML for the envelope model are also limited to the transverse directions for the moment.
An extension to the method along the longitudinal direction may prove useful in order to support an even larger range of applications.

\section*{Acknowledgements}

This work was granted access to the HPC resources of IDRIS under the allocation 2021-A0090510062 (Virtual Laplace) made by GENCI. Most of the development was made on the meso-scale HPC "3Lab Computing" hosted at \'{E}cole polytechnique and administrated by the Laboratoire Leprince-Ringuet, Laboratoire des Solides Irradiés et Laboratoire pour l'Utilisation des Lasers Intenses.
The authors would like to thank Julien Derouillat who kickstarted the implementation of the PML in \Smilei.

\bibliographystyle{unsrt}
\bibliography{Bibliography}
\end{document}